\documentclass{emulateapj}
\RequirePackage{natbib}
\usepackage{latexsym}

\begin{document}

\title{MAGNETIC BRAKING AND VISCOUS DAMPING OF DIFFERENTIAL ROTATION IN
       CYLINDRICAL STARS}
\shorttitle{MAGNETIC BRAKING OF DIFFERENTIAL ROTATION}

\author{James N. Cook, Stuart L. Shapiro\altaffilmark{1}, and Branson C. Stephens}

\affil{Department of Physics, University of Illinois at
        Urbana-Champaign, Urbana, Il 61801}

\altaffiltext{1}{Department of Astronomy and NCSA, University of Illinois at
        Urbana-Champaign, Urbana, Il 61801}

\begin{abstract}
Differential rotation in stars generates toroidal 
magnetic fields whenever an initial seed poloidal field is present. 
The resulting magnetic stresses, along with viscosity, drive the star 
toward uniform rotation.  This magnetic braking has important 
dynamical consequences in many astrophysical contexts.  For example, 
merging binary neutron stars can form ``hypermassive'' remnants 
supported against collapse by differential rotation.  The removal of 
this support by magnetic braking induces radial fluid motion, which can 
lead to delayed collapse of the remnant to a black hole.  We explore the 
effects of magnetic braking and viscosity on the structure of a 
differentially rotating, compressible star, generalizing our earlier 
calculations for incompressible configurations.  The star is idealized 
as a differentially rotating, infinite cylinder supported initially by a 
polytropic equation of state.  The gas is assumed to be infinitely 
conducting and our calculations are performed in Newtonian gravitation.  
Though highly idealized, our model allows for the incorporation of 
magnetic fields, viscosity, compressibility, and shocks with minimal 
computational resources in a 1+1 dimensional Lagrangian MHD code.  
Our evolution calculations show that magnetic braking can lead to 
significant structural changes in a star, including quasistatic 
contraction of the core and ejection of matter in the outermost 
regions to form a wind or an ambient disk. These calculations serve 
as a prelude and a guide to more realistic MHD simulations in full 
3+1 general~relativity.
\end{abstract}


\section{INTRODUCTION}
Magnetic fields play a crucial role in determining the evolution of
many relativistic objects.  Some of these systems are promising 
sources of gravitational radiation for detection by laser 
interferometers now under design and construction, such as LIGO,
VIRGO, TAMA, GEO, and LISA.  For example, merging neutron stars 
can form a differentially 
rotating ``hypermassive'' remnant (Rasio \& Shapiro 1992, 1999; 
Baumgarte, Shapiro, \& Shibata 2000; Shibata \& Ury\=u 2000).  These 
configurations have rest 
masses that exceed both the maximum rest mass of nonrotating spherical 
stars (the TOV limit) and uniformly rotating stars at the mass-shedding 
limit (``supramassive stars''), all with the same polytropic index. 
This is possible because differentially rotating neutron stars can 
support significantly more rest mass than their nonrotating or 
uniformly rotating counterparts. Baumgarte, Shapiro \& Shibata (2000), 
have performed 
dynamical simulations in full general relativity to demonstrate that 
hypermassive stars can be constructed that are dynamically stable 
against radial collapse and nonradial bar formation.  The dynamical 
stabilization of a hypermassive remnant by differential rotation may 
lead to delayed collapse and a delayed gravitational wave burst.  
The reason is that the stabilization due to differential rotation, 
although expected to last for many dynamical timescales (i.e. many 
milliseconds), will ultimately be destroyed by magnetic braking and/or 
viscosity. These processes drive the star to uniform rotation, which 
cannot support the excess mass, and will lead to catastrophic collapse, 
possibly accompanied by some mass~loss.

\pagebreak

Baumgarte \& Shapiro (2003) discuss several other relativistic 
astrophysical 
scenarios in which magnetic fields will be important.  Neutron stars
formed through core collapse supernovae are probably characterized by
significant differential rotation (see, {\em e.g.}, Zwerger \& 
M\"{u}ller 1997; Rampp, M\"{u}ller, \& Ruffert 1998; Liu \& Lindblom 2001; 
Liu 2002, and references therein).
Conservation of angular momentum during the collapse is expected to
result in a large value of $\beta = T/|W|$, where $T$ is the 
rotational kinetic energy and $W$ is the gravitational potential 
energy.  However, uniform rotation can 
only support small values of $\beta$ 
without leading to mass shedding (Shapiro \& Teukolsky, 1983).  Thus, 
nascent neutron stars from supernovae probably rotate differentially,  
giving magnetic braking and viscosity an important dynamical
role.  Short period gamma-ray bursts (GRB's) are 
thought to result from binary neutron star mergers (Narayan, 
Paczynski, \& Piran 1992) or tidal disruptions of neutron stars 
by black holes (Ruffert \& Janka 1999).  Meanwhile, long period
GRB's are believed to result from collapses of 
rotating massive stars to form black holes (MacFadyen \& Woosley 1999).
In most current models of GRB's, 
the burst is powered by rotational energy extracted from the neutron 
star, black hole, or surrounding disk (Vlahakis \& Konigl 2001).  
This energy extraction can 
be accomplished by strong magnetic fields, which also provide an 
explanation for the collimation of GRB outflows into jets
(Meszaros \& Rees 1997; Sari, Piran, \& Halpern 1999; Piran 2002) and 
for the observed gamma-ray polarization (Coburn \& Boggs 2003).  
The 
dynamics of supermassive stars (SMS's), which may have been present 
in the early universe, will also depend on magnetic fields if the 
SMS's rotate differentially.  Loss of differential rotation support 
will affect the collapse of SMS's, which has been proposed as a 
formation mechanism for supermassive black holes observed in galaxies 
and quasars, or their seeds (see Rees 1984, Baumgarte \& Shapiro 1999,
Bromm \& Loeb 2003, and Shapiro 2003 for 
discussion and references).  Finally, the effectiveness of the $r$-mode 
instability 
in rotating neutron stars may depend on magnetic fields.  This 
instability has been proposed as a mechanism for limiting the angular 
velocities of neutron stars and for producing quasi-periodic 
gravitational waves (Andersson 1998; Friedman \& Morsink 1998;
Andersson, Kokkotas, \& Stergioulas 1999; Lindblom, Owen, \&
Morsink 1998).  Due to flux-freezing, however, the magnetic 
fields may distort or suppress the $r$-modes (see Rezzolla et al.\  2000,
2001a,b and references~therein). 

Motivated in part by the growing list of important, unsolved problems
which involve hydromagnetic effects in strong-field dynamical 
spacetimes, Shapiro (2000, hereafter Paper I) performed a simple,
Newtonian, MHD calculation to highlight the importance of magnetic 
fields in differentially rotating stars.  In this model, the star 
is idealized as a differentially 
rotating infinite cylinder consisting of a homogeneous, incompressible 
conducting gas in hydrostatic equilibrium. The magnetic field is taken 
to be purely radial initially and 
is allowed to evolve according to the ideal MHD equations.  
A constant shear viscosity is also 
allowed.  Note that, though highly idealized, this model allows 
several important physical effects to be included in a 1+1 
evolution code, requiring much less computational effort
than more realistic 3+1 \linebreak  simulations.  

Paper I presents results for the behavior of differentially rotating 
cylinders in three physical 
situations: zero viscosity with a vacuum exterior, nonzero viscosity
with a vacuum exterior, and zero viscosity with a tenuous ambient 
plasma atmosphere.  
In the first case, for which the solution is analytic, the magnetic 
field gains a toroidal component which oscillates back and forth in
a standing Alfv\'en wave pattern.  The angular velocity profile
oscillates around a state of uniform rotation, with uniform rotation
occurring at times when the azimuthal magnetic field is at its
maximum magnitude.  At these times, a significant amount of the 
rotational energy has been converted to magnetic energy.  In the 
absence of a dissipation mechanism, these oscillations continue 
indefinitely. In the second case (vacuum exterior with nonzero 
viscosity), the oscillations are damped and the cylinder is driven 
to a permanent state of uniform rotation with a significant fraction 
of its initial rotational energy having been converted into magnetic 
energy and finally into heat.  In the final case (zero viscosity 
with a plasma atmosphere), Alfv\'en waves are partially transmitted 
at the surface and carry away significant fractions of the angular 
momentum and energy of the cylinder.  In all of these situations, only 
the timescale, and not the qualitative outcome, of the dynamical 
behavior depends on the strength of 
the initial magnetic field.  Thus, even for a small initial seed 
field, magnetic braking and viscosity eventually drive the cylinders 
toward uniform~rotation. 

The braking of differential rotation by magnetic fields is also 
currently being studied in a spherical, incompressible neutron star 
model (Liu \& Shapiro, in preparation).  This calculation
treats the slow-rotation, weak-magnetic field 
case in which $T \ll E_{\rm mag} \ll |W|$, where $E_{\rm mag}$ is
the total magnetic energy. Consequently, the star
is spherical to a good approximation and poloidal velocities in the $\theta-$
and $r-$ directions can be neglected on the timescales of interest.
Liu \& Shapiro solve the MHD equations for the coupled evolution 
of the angular velocity
$\Omega$ and the azimuthal magnetic field $B_{\phi}$ in both
Newtonian gravity and general relativity.  The resulting angular 
velocity and poloidal field 
profiles often develop rich small-scale structure due to 
the fact that the eigenfrequencies of $B_{\phi}$ vary with location
inside the~star.
    
In many cases, the loss of differential rotation support due to 
magnetic braking is expected to lead to interesting dynamical 
behavior.  In hypermassive stars, magnetic braking may lead to 
catastrophic gravitational collapse.  In general, however, one
expects a range of radial motions, including oscillations, 
quasistatic contraction, and ejection of winds.  These behaviors
are not present in the results of Paper I because of the assumptions
of homogeneity and incompressibility.  In that case, no radial 
velocities occur as the system evolves away from equilibrium.
In the present paper, we explore the effects of compressibility 
by generalizing the calculations of Paper I to Newtonian cylindrical 
polytropes.  Though other aspects of the present model are again 
highly idealized, allowing for compressible matter results in a 
much more realistic spectrum of dynamical behavior, including radial 
contraction, mass ejection, and shocks.  By formulating the problem 
in 1+1 dimensions and working in Lagrangian coordinates, we are 
able to solve the MHD equations essentially to arbitrary precision.
Our simulations again serve to identify most of the key physical and 
numerical parameters, scaling behavior, and competing timescales 
associated with magnetic braking and differential rotation. The 
structure of this paper is as follows: Section~\ref{basicmodel} 
describes our basic model.  In Section~\ref{eqbm}, we discuss the 
structure of equilibrium cylindrical polytropes which serve as 
initial data for our evolution calculations.  In 
Section~\ref{mhdeqns}, we set out the fundamental MHD evolution 
equations and put them in a convenient nondimensional form.    
Section~\ref{results} describes the results of MHD evolution
calculations for several choices of the parameters which serve to 
illustrate the effects of compressibility and differential rotation. 
We discuss the significance of these calculations in 
Section~\ref{conclusions}.

\section{BASIC MODEL}
\label{basicmodel}
As discussed in Paper I, differential rotation in a spinning star 
twists up a frozen-in, poloidal magnetic field, creating
a very strong toroidal field. This process will generate Alfv\'{e}n 
waves, which can redistribute the angular momentum 
of a star on timescales less than $\sim~100\,{\rm s}\,(10^{12} {\rm G}/B_0)$,  
where $B_0$ is the characteristic initial poloidal field. Shear 
viscosity also redistributes angular momentum. However, molecular 
viscosity in neutron star matter operates on a typical timescale of 
$\sim~10^9{\rm s}$ in a star of $\sim~10^9{\rm K}$, so it 
alone is much less effective in bringing the star 
into uniform rotation than magnetic braking, unless the initial 
magnetic field is particularly weak. Turbulent viscosity, 
if it arises, can act more~quickly. 

We wish to track the evolution of differential rotation and 
follow the competition between magnetic braking and viscous 
damping.  Following Paper I,  we model a spinning star as 
an infinite, axisymmetric cylinder with a vacuum exterior. We adopt 
cylindrical coordinates $(r,\phi,z)$, with the $z-$ axis aligned 
with the rotation axis of the star, and assume translational symmetry 
in the $z-$ direction. We take the magnetic field 
to have components only in the $r-$ and $\phi-$ directions.
The fluid is assumed to be perfectly conducting everywhere
(the ideal MHD regime).  In some of our models, 
we allow for the presence of a constant shear viscosity.  
For protoneutron stars with $T \gtrsim 10^9 K$, the coefficient
of bulk viscosity, due to the time lag for reestablishing beta
equilibrium following a change in density, becomes comparable
to the shear viscosity coefficient (Sawyer 1989).  We do not
account for bulk viscosity, however, since our models are strongly
differentially rotating (and hence have large amounts of shear), while
large scale density changes in our evolution calculations occur 
quasistatically (see Section~\ref{results}).  

To construct initial data for our dynamical simulations, we 
assume a polytropic equation of state (EOS): $P = K\rho^{\Gamma}$. 
The equations of hydrostatic equilibrium are then solved with an 
assumed form for the initial differential rotation law (discussed
in Section~\ref{eqbm}).  We take the $\phi-$ component of the 
magnetic field to be zero initially. We then evolve the system by 
numerically integrating the full set of coupled MHD equations.  
We allow for heating due to shocks and viscous dissipation
during the evolution, and assume a $\Gamma-$law EOS, 
$P=(\Gamma-1)\rho\varepsilon$, where $\varepsilon$ is the specific
internal~energy.

\section{EQUILIBRIUM CYLINDRICAL POLYTROPES}
\label{eqbm}

Here we discuss the initial data for our simulations.
We derive the equations for the static case, and then generalize to include
rotation and radial magnetic fields.  Finally, we derive a virial relation 
for equilibrium infinite cylindrical stars, discuss the stability of 
these models to radial perturbations, and construct some numerical~models.

\subsection{\em Static Polytropes}

The fundamental equations for a static polytrope are the polytropic~EOS
\begin{equation}
P = K\rho^\Gamma = K\rho^{1+\frac{1}{n}},      \label{poly}
\end{equation}
where $\rho$ is the density, $P$ is the pressure, $K$ and $\Gamma$ are constants, and
$n$ is the polytropic index, and the equation of hydrostatic~equilibrium
\begin{equation}
\frac{1}{\rho}{\mathbf \nabla}P = -{\mathbf \nabla}\Phi,    \label{HydroEq}
\end{equation}
where $\Phi$ is the Newtonian gravitational potential, which satisfies Poisson's~equation,
\begin{equation}
\nabla^2 \Phi = 4\pi\rho.      \label{Poisson}
\end{equation}
Note that we will work in units such that $G=1$.   
To find ${\mathbf \nabla}\Phi$, we first define $\mu(r)$ to be the  mass per unit 
length interior to radius $r$.  This~gives
\begin{equation}
\mu(r)= 2\pi\int_0^r\rho(r')r'dr'.  \label{mudef}
\end{equation}
Integrating equation (\ref{Poisson}) and imposing regularity at the origin~yields
\begin{equation}
\frac{d\Phi}{dr} = \frac{2\mu(r)}{r}.     \label{gradphi}
\end{equation}
Differentiating equation~(\ref{HydroEq}) in cylindrical radius $r$ and 
combining with equation~(\ref{Poisson})~yields 
\begin{equation}
\frac{d}{dr}\left(\frac{r}{\rho}\frac{dP}{dr}\right) = -4\pi\rho(r)r. \label{Hydrofinal}
\end{equation}
This equation can be generalized for infinite planar or 
spherical geometries through similar arguments.  The result~is
\begin{equation}
\frac{d}{dr}\left(\frac{r^{\alpha-1}}{\rho}\frac{dP}{dr}\right) = 
-4\pi\rho(r)r^{\alpha-1}, \label{GenHydro}
\end{equation}
where $\alpha = 1,2,3$ corresponds to planar, cylindrical, and spherical 
geometry~respectively.  

We will make use of a common nondimensionalization of equation 
(\ref{GenHydro}) to
facilitate numerical solutions.  We define nondimensional quantities
according~to
\begin{eqnarray}
\rho &=& \rho_c \theta^n                       \nonumber  \\
r &=& a \xi                                    \label{lenondim}      \\
a &=& \left[ \frac{(n+1)K\rho_c^{(1/n-1)}}{4\pi}\right]^{1/2},  \nonumber
\end{eqnarray}
where $\rho_c$ is the central density. Equation (\ref{GenHydro}) then
becomes a {\it generalized} Lane-Emden~equation,
\begin{equation}
\frac{1}{\xi^{\alpha-1}}\frac{d}{d\xi}\left(\xi^{\alpha-1}\frac{d\theta}{d\xi}
\right) = -\theta^n.   \label{lane}
\end{equation}
Hereafter we will treat $\alpha = 2$ (cylindrical geometry).  Spherical 
polytropes were described extensively by Chandrasekhar (1939); planar 
polytropes were constructed by Shapiro (1980).  The boundary conditions on 
equation (\ref{lane}) are easily obtained from physical arguments.  From 
equation (\ref{lenondim}), it is evident that $\theta(0) = 1$.  The 
condition on $\theta'$ is obtained by using the fact that $\rho \to \rho_c$ 
so that equation (\ref{mudef}) implies $\mu(r) \propto r^2$ as $r \to 0$, hence 
equations (\ref{poly}), (\ref{HydroEq}), and (\ref{gradphi}) yield 
$d\rho/dr = 0$, and hence $\theta'(0) = 0$.  The equations should be 
integrated from the origin out until $\theta = 0$, which implies $P = 0$.  
This point is the surface of the star and will be denoted by $\xi_1$.  
This gives the pressure and density profiles for equilibrium configurations.  
Some analytic solutions to equation (\ref{lane}) are, for $n=0$,
$\theta = 1-\xi^2/4$ and for $n=1$, $\theta = J_0(\xi)$, where $J_0$ is the
Bessel function of order~zero.  

\subsection{\em Rotating Cylindrical Polytropes}

We will now extend the cylindrical polytrope model to include the effects 
of rotation.  The hydrostatic equilibrium equation is modified by a term 
accounting for centrifugal acceleration.  The result~is
\begin{equation}
\frac{1}{\rho}\frac{dP}{dr} = -\frac{d\Phi}{dr}+\Omega^2(r)r,    \label{Hydrostead}
\end{equation}
where $\Omega(r)$ is the angular velocity at radius $r$ (the $r$-dependence 
allows for differential rotation).  Performing the same manipulations used 
to get equation~(\ref{Hydrofinal})~gives
\begin{equation}
\frac{d}{dr}\left(\frac{r}{\rho}\frac{dP}{dr}\right)= 2\Omega^2r + 
r^2\frac{d\Omega^2}{dr}-4\pi\rho r.  \label{HydroRot}
\end{equation}
Using the same nondimensionalization as equation (\ref{lenondim}) and making 
the~definition
\begin{equation}
\zeta \equiv \frac{\Omega^2}{\pi\rho_c},
\label{zetadef}
\end{equation}
gives the cylindrical Lane-Emden equation with~rotation,
\begin{equation}
\frac{d^2\theta}{d\xi^2}+\frac{1}{\xi}\frac{d\theta}{d\xi}+\theta^n 
- \frac{\zeta}{2} -\frac{\xi}{4}\frac{d\zeta}{d\xi}= 0.\label{rotLE}
\end{equation}
By identical reasoning, this equation has the same boundary conditions as 
equation~(\ref{lane}).

There are two analytic solutions for uniform rotation ($\zeta = {\rm constant}$).  
The analytic solutions are for $n=0$, $\theta(\xi) = 1 - (1-\zeta/2)\xi^2/4$ and 
for $n=1$, $\theta(\xi) = (1-\zeta/2)J_0(\xi)+\zeta/2$.  In the rest of the paper, 
we will adopt the following $\Omega$~profile
\begin{equation}
\Omega(r) = \frac{\Omega_0}{2} \left[ 1+\cos \left(\frac{\pi r^2}{R^2} \right) \right],
\label{rotlaw}
\end{equation}
where $R$ is the radius of the~star.
Letting $\zeta_0 = \Omega_0^2/\pi\rho_c$ gives the nondimensional~form
\begin{equation}
\zeta(\xi)= \frac{\zeta_0}{4} \left[ 1+\cos \left(\frac{\pi \xi^2}{\xi_1^2} \right) 
\right]^2. \label{nonrot}
\end{equation}
Though this rotation law was chosen arbitrarily, it satisfies the boundary 
condition given below (eq. [\ref{domegabound}]) and has the expected property that
$\Omega$ decreases monotonically with $r$.  Shibata and Ury\=u (2000) found that
hypermassive binary merger remnants typically have $\Omega(0)/\Omega(R) 
\approx 3$.  Slightly modifying our rotation law to satisfy this condition 
resulted in a set of models qualitatively similar to those discussed below, 
usually with slightly larger allowed values of $\beta = T/|W|$.  
To solve equation (\ref{rotLE}) for rotation profile of equation (\ref{nonrot}), 
we must iterate our solutions to equation (\ref{rotLE}) to identify
$\xi_1$, the surface radius of our star. As will be shown later, these 
equilibrium solutions are also valid for cylinders with radial magnetic fields, 
so we will use these solutions as initial data for our dynamical evolutions.  We
remark how simple it is to construct differentially rotating, infinite cylindrical 
polytropes, described by ordinary differential equations in one dimension, in 
comparison to finite rotating stars, which are defined by partial differential 
equations in two dimensions (see {\it e.g.} Tassoul~1978).

\subsection{\em Virial Theorem and Stability}
\label{virsection}

We will use the virial theorem to check the results of our equilibrium 
calculations and, later on, to test whether a configuration has reached 
a new equilibrium state following dynamical evolution.  With 
the definitions
\begin{eqnarray}
W & \equiv & - \int \rho {\mathbf x} \cdot {\mathbf \nabla} \Phi d^3x, \nonumber \\
T & \equiv & \frac{1}{2}\int \rho \upsilon^2 d^3x,  \label{ints} \\
\Pi & \equiv & \int P d^3x,  \nonumber  
\end{eqnarray}
the virial theorem for an infinite cylindrical equilibrium star~is
\begin{equation}
2T + W + 2\Pi = 0.  \label{vir}
\end{equation}
Notice that the numerical coefficients in this equation differ from the 
analogue for bounded configurations.  (For a derivation, see 
Appendix~\ref{virial}.)  Also note that as shown in Appendix \ref{virial}, 
even if magnetic fields are present, this equation still holds unmodified 
for equilibrium stars.  In our case it is convenient to write
$d^3x = L d\mathcal{A} = 2\pi Lrdr$, where $L$ is an arbitrary length in 
the $z$-direction.  We divide each quantity in equation (\ref{ints}) by $L$ 
to obtain quantities per unit length (we will continue to use the symbols 
$W$, $T$, and $\Pi$).  Hereafter, we will refer to all extensive quantities 
as quantities per unit length in the $z-$ direction.  These definitions give 
an identical virial theorem to equation (\ref{vir}).  Using equations 
(\ref{mudef}) and (\ref{gradphi}), the definition of $W$ can be integrated 
analytically to~give
\begin{equation}
W = -\mu_t^2,  \label{Wint}
\end{equation}
where $\mu_t = \mu(R)$ is the total mass per unit length of the~star.

The virial theorem can be used to find an upper limit on $\beta = T/|W|$ 
for any equilibrium cylindrical star.  Since the pressure is always 
positive, $\Pi$ is positive definite.  Similarly, $\mu_t^2$ is positive
definite, so $W$ is negative or $W = -|W|$.  Using these facts in 
equation~(\ref{vir}) implies that for any equilibrium~star
\begin{equation}
\beta \leq \frac{1}{2}.            \label{blim}
\end{equation}

We require our equilibrium solutions to be initially stable to radial 
perturbations to ensure that we are seeing the secular effects of magnetic 
fields in the evolution and not a hydrodynamical radial instability.  For 
cylinders, the critical $\Gamma$ for stability is $\Gamma \geq 1$, 
permitting any $n \geq 0$ (see Appendix \ref{stability}).  For spheres,
however, one needs $\Gamma \geq 4/3$, which in turn requires $n \leq 3$.  
Note that rotation lowers the critical $\Gamma$, which does not change 
the limit on $n$, since any realistic EOS allows us to restrict ourselves to 
$n \geq 0$.  Also note that the presence of a radial magnetic field does 
not affect this stability analysis, as shown in Appendix~\ref{stability}.

\subsection{\em Equilibrium Models}

Here we present the results of our equilibrium model calculations.  We 
write equation (\ref{rotLE}) as a system of two coupled first order ODE's 
and integrate using a fourth order Runge-Kutta method.  Below we discuss 
the comparison with an {\em analytic} Roche model for centrally condensed rotating 
cylinders and give a table of data for cylindrical~polytropes.

\subsubsection{\em Roche Model}

The Roche model holds for uniformly rotating, centrally condensed, 
polytropic cylinders.  Our discussion is patterned after Shapiro \& Teukolsky
(1983). We will show below that this approximate model improves with 
increasing $n$, as expected.  Equation (\ref{Hydrostead}) may be written~as
\begin{equation} 
\frac{1}{\rho}\nabla P + \nabla(\Phi + \Phi_c) = 0,   \label{HydroRoche}
\end{equation}
where $\Phi_c = -\Omega^2 r^2/2$.  Because of the central condensation, 
near the surface $\mu \to \mu_t$~and 
\begin{equation}
\frac{\partial \Phi}{\partial r} \simeq \frac{2\mu_{t}}{r} \Rightarrow 
\Phi \simeq 2\mu_{t}\ln r.
\end{equation}
The specific enthalpy of a polytrope is given~by
\begin{equation}
h(r) = \int^r dr' \frac{1}{\rho} \frac{\partial P}{\partial r'} =
\frac{\Gamma}{\Gamma - 1}\frac{P}{\rho}.
\end{equation}
Equation (\ref{HydroRoche}) can now be integrated to~obtain
\begin{equation}
h(r) + \Phi(r) + \Phi_c(r) = K,    \label{moment}
\end{equation}
where $K$ is an integration constant.  Since the star is very centrally
condensed and equation (\ref{moment}) holds throughout the envelope, $K$ 
is assumed to be the same as in the nonrotating case.  Let $R_0$ be the 
radius of the nonrotating star.  Then, since the enthalpy vanishes at 
the~surface,
\begin{equation}
\Phi(R_0)=K = 2\mu_t\ln R_0.
\end{equation}

A stable rotating star ({\em i.e.}\ not shedding mass) must have a radius 
$R_1$ such that $h(R_1)=0$.  Let the effective potential be defined as
$\Phi_{\rm eff}=\Phi+\Phi_c$ and let the maximum of the effective potential
be $\Phi_{\rm max}$.  Then using equation (\ref{moment}) and the fact that 
$h(r)$ has a zero, this stability condition~becomes 
\begin{equation}
\Phi_{\rm max} \geqslant K,
\end{equation}
where $\Phi_{\rm max}$ is determined by setting $\partial \Phi_{\rm eff}/\partial
r = 0$,~giving 
\begin{equation}
r = \sqrt\frac{2\mu_t}{\Omega^2}.
\end{equation}
Inserting this value into $\Phi_{\rm max}$ gives the rotational stability~condition
\begin{equation}
-\mu_t + 2\mu_t \ln \sqrt\frac{2\mu_t}{\Omega^2} \geqslant
2\mu_t\ln R_0. 
\end{equation}
This implies that to avoid mass shedding, the star must~satisfy 
\begin{equation}
\Omega^2 \leqslant \frac{2}{e}\frac{\mu_t}{R_0^2}
\label{omegacond}
\end{equation}
Then in terms of nondimensional variables of the nonrotating star, the 
stability criterion in the Roche approximation~becomes
\begin{equation}
\zeta \leqslant -\frac{1}{e} \frac{4}{\xi_1} \left. 
\frac{\partial \theta}{\partial \xi}\right|_{\xi=\xi_1},
\label{zetacond}
\end{equation}
employing equations (\ref{gradphi}), (\ref{zetadef}), and (\ref{omegacond}).
Now, we use our code to find the maximum $\zeta$ for solutions to equation 
(\ref{rotLE}) and compare to the Roche model results.  We denote the maximum 
$\zeta$ from our code as $\zeta_m$ and the maximum $\zeta$ from the Roche model, 
equation~(\ref{zetacond}), as $\zeta_R$. We find the following sets 
$(n,\zeta_m,\zeta_R)$: (1.0, 0.57425, 0.31767), (3.0, 0.10948, 0.08527), 
(5.0, 0.02977, 0.02659), and (10.0, 0.001795, 0.001767).  These results confirm
that the Roche model becomes increasingly more accurate as $n$ increases and
the degree of central concentration~increases.

\subsubsection{{\em Numerical Results}}

\begin{table}
\begin{center}
\centerline{\sc Table 1}
\centerline{\sc Static Cylindrical Polytropes}
\vskip 6pt
\begin{tabular}{c c c c}
\hline 
\hline \vspace{0.03in}
$n$ & $\xi_1$ & 
$-\xi_1\left(\frac{d\theta}{d\xi}\right)_{\xi_1}$ & 
$\rho_c/\bar{\rho}$ \\ 
\hline 
0.0  & 2.0    & 2.0   & 1.0    \\
1.0  & 2.405  & 1.248 & 2.316  \\
2.0  & 2.921  & 0.925 & 4.611  \\
3.0  & 3.574  & 0.740 & 8.629  \\
5.0  & 5.428  & 0.532 & 27.67  \\ 
\hline
\end{tabular}
\end{center}
\end{table}

\begin{figure}
\plotone{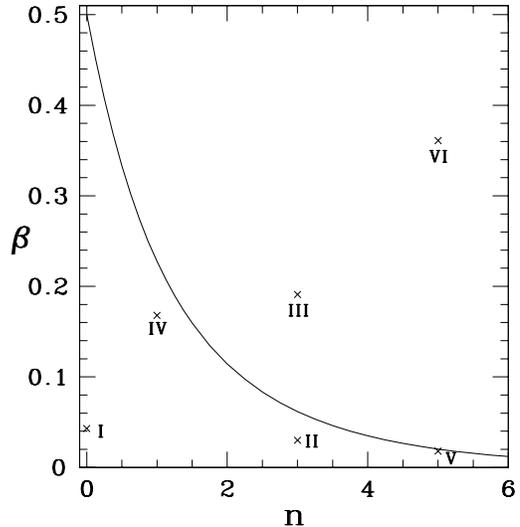}
\caption{Maximum $\beta = T/|W|$ attainable for uniformly 
  rotating stars (mass shedding limit; the solid line) and the 
  values of $\beta$ which we will evolve numerically (marked 
  with crosses and labeled I-VI). Note that the virial theorem 
  sets the limit $\beta \leq 0.5$ for any equilibrium~configuration.
  \label{twplot}}
\end{figure}

Table~1 summarizes numerical results for static cylindrical polytropes.  As 
$n$ increases, the star becomes more centrally condensed.  However, this is 
much less pronounced than in the spherical case, where $n=3$ gives 
$\rho_c/\bar{\rho} = 54.18$ compared to $\rho_c/\bar{\rho} = 8.629$ for 
cylinders. Table~2 gives the key results for uniformly rotating cylindrical 
polytropes spinning at the mass-shedding limit.  Note that we have not found 
a maximum $n$ for solutions with finite $\mu_t$ in the case of cylindrical 
polytropes.  In contrast, polytropic spheres with $n > 5$ extend to infinite 
radius with infinite mass (Chandrasekhar 1939).  An important consequence of 
our differential rotation law is that for $n \gtrsim 1.2$, we can get a 
larger value of $\beta$ than the maximum $\beta$ for uniform rotation, 
$\beta_{\rm max}$.  Since we expect viscosity and magnetic braking to drive 
differential rotation toward uniform rotation, we expect the most interesting 
cases to be those where the star is forbidden to simply relax to uniform 
rotation.  Figure~\ref{twplot} shows a plot $\beta_{\rm max}$ versus $n$, 
where we have marked the cases we dynamically evolve below after  
incorporating magnetic fields and viscosity.  We expect the cases falling 
below the curve to relax to a uniformly rotating equilibrium polytrope and 
the cases above the curve to undergo major changes away from 
polytropic~behavior.

\begin{table*}
\begin{center}
\centerline{\sc Table 2}
\centerline{\sc Maximally Rotating Cylindrical Polytropes}
\vskip 6pt
\begin{tabular}{c c c c c}
\hline
\hline
  & \multicolumn{2}{c}{Uniform Rotation\tablenotemark{a}} & 
\multicolumn{2}{c}{Differential Rotation\tablenotemark{b}} \\
n & $\beta_{max}$ &  
\small{$\Omega_{max}/\sqrt{\pi\rho_c}$}
& $\beta_{max}$ & 
\small{$\Omega_{max}/\sqrt{\pi\rho_c}$} \\
\hline
0.0   & 0.500 & 1.414 & 0.086 & 1.414  \\
0.01  & 0.496 & 1.402 & 0.087 & 1.414  \\ 
1.0   & 0.228 & 0.758 & 0.179 & 1.414  \\   
3.0   & 0.062 & 0.331 & 0.299 & 1.414  \\
5.0   & 0.015 & 0.172 & 0.368 & 0.382  \\
\hline
\hline
\end{tabular}
\vskip 12pt
\begin{minipage}{10cm}
${}^{a}$ {The maximum corresponds to the mass shedding limit.}
\\
${}^{b}$ {Assumes the particular rotation law 
$\Omega~=~\frac{\Omega_0}{2}\left[1+\cos\left(\frac{\pi r^2}{R^2}\right)\right]$ 
and a monotonically decreasing density profile.}
\end{minipage}
\end{center}
\end{table*}

\section{MHD EQUATIONS}
\label{mhdeqns}
We now assemble the fundamental equations for our MHD
simulations.  First, the fluid obeys the continuity~equation,
\begin{equation}
\frac{\partial\rho}{\partial t} + {\mathbf \nabla}\cdot(\rho{\mathbf v}) = 0.
\label{continuity}
\end{equation}
The fluid motion is governed by the magnetic Navier-Stokes equation:
\begin{eqnarray}
\frac{\partial{\mathbf v}}{\partial t} + 
({\mathbf v}\cdot{\mathbf \nabla}){\mathbf v} & = & 
-\frac{1}{\rho}{\mathbf\nabla}P - {\mathbf \nabla}\Phi +
\frac{{(\mathbf \nabla}\times{\mathbf B})\times{\mathbf B}}{4\pi\rho} 
\nonumber \\ 
& & + \frac{\eta}{\rho}(\nabla^2{\mathbf v} + 
\onethird{\mathbf\nabla}({\mathbf\nabla}\cdot{\mathbf v})),
\label{navier}
\end{eqnarray}
where $\mathbf{B}$ is the magnetic field and $\eta$ is the coefficient of 
dynamic viscosity, which is related to the familiar kinematic viscosity 
according to $\nu = \eta/\rho$.  
We note that if ${\mathbf B}=0=\eta$, and $\mathbf{v}=\mathbf{\Omega}
\times\mathbf{r}$, $\mathbf{\Omega}=\Omega \mathbf{\hat{z}}$, 
equation (\ref{navier}) reduces to the hydrostatic equilibrium 
equation for rotating cylinders (eq.\ [\ref{Hydrostead}]).  The magnetic 
field $\mathbf{B}$ satisfies Maxwell's constraint~equation
\begin{equation}
\mathbf{\nabla}\cdot \mathbf{B}\ =\ 0,
\label{divb}
\end{equation}
as well as the flux-freezing~equation
\begin{equation}
{\partial \mathbf{B} \over \partial t}\ 
=\ \mathbf{\nabla}\times(\mathbf{v}\times \mathbf{B}).
\label{fluxfreeze}
\end{equation}

Solving equation (\ref{divb}) for the magnetic field 
together with the flux-freezing condition (\ref{fluxfreeze}) 
requires that the radial component of the magnetic field be 
independent of time and given~by
\begin{equation}
B_r = {B_0 R \over r}, ~~~t \geq 0,
\label{Br}
\end{equation}
where $B_0$ is the value of the field at the stellar surface at $t=0$. 
Although the magnetic field given by equation (\ref{Br}) exhibits a 
static line singularity along the axis at $r=0$, it does not drive 
singular behavior in the fluid velocity or nonradial magnetic field. 
As these quantities, which are the main focus here, remain finite 
and evolve in a physically reasonable fashion, the line singularity 
poses no difficulty and requires no special treatment.  Also, since
the magnetic term in equation (eq.\ [\ref{navier}]) is proportional to
the curl of $\mathbf{B}$, the radial magnetic field does not enter
the hydrostatic equilibrium equation.  Thus, the presence of an 
initially radial magnetic field does not affect the initial data 
equilibrium models described in Section~\ref{eqbm}.  The virial 
theorem derived by taking a moment of equation (\ref{navier}) is 
identical to equation (\ref{vir}), which applies in the 
${\mathbf B} = 0$ case.  (The magnetic field terms cancel as described  
in Appendix \ref{virial}.)  Note that the virial theorem only holds 
for equilibrium configurations and is not expected to be satisfied 
during a dynamical~evolution.
   
We next consider the properties of the fluid itself.  The equation of 
state, $P=(\Gamma-1)\rho\varepsilon$, allows 
the entropy of a fluid element to increase if there is heating
due to shocks or shear viscosity.  The evolution of the specific
internal energy is described by the First Law of~Thermodynamics,
\begin{equation}
\frac{d\varepsilon}{dt} = -P\frac{d}{dt}\left(\frac{1}{\rho}\right)
+T \frac{ds}{dt},
\label{einteq}
\end{equation} 
where the Lagrangian derivative is defined as $d/dt \equiv \partial/\partial t
+ {\mathbf v}\cdot\nabla$ and $s$ is the specific entropy.  The rate of energy 
generation due to viscosity is given~by\footnote{Equations (\ref{eintvisc}) 
and (\ref{visctensor}) are expressed in Cartesian coordinates for~simplicity.}
\begin{equation}
T \frac{ds}{dt} = 
\frac{1}{\rho}\sigma'_{ij}\frac{\partial \upsilon_i}{\partial x_j},
\label{eintvisc}
\end{equation}
where $\sigma'_{ij}$ is the viscous stress tensor (Landau \& Lifshitz 1998, eq.\ 15.3)
\begin{equation}
\sigma'_{ij} = \eta \left(\frac{\partial\upsilon_i}{\partial x_j}
+\frac{\partial\upsilon_j}{\partial x_i} - \twothirds \delta_{ij}
\frac{\partial\upsilon_k}{\partial x_k}\right),
\label{visctensor}
\end{equation}
and where summation over repeated indices is~implied. 

We will now write our system of evolution equations in component
form in Lagrangian coordinates (we work in an orthonormal basis).  
First, we define $j$ as the specific angular~momentum: 
$j \equiv r\upsilon_{\phi}$, where $\upsilon_{\phi}$ is
the $\phi-$ component of the velocity.  We also explicitly use the  
gradient of $\Phi$ given in equation (\ref{gradphi}).  Then 
equations (\ref{continuity}), (\ref{navier}), (\ref{fluxfreeze}), 
(\ref{einteq}), and (\ref{eintvisc})~become
\begin{eqnarray}
\label{densevolve}
\frac{d\rho}{dt} & = & - \frac{\rho}{r}\frac{\partial}{\partial r}(r\upsilon_r) \\
\frac{d\upsilon_r}{dt} & = & -\frac{1}{\rho}\frac{\partial P}{\partial r} 
 - \frac{2\mu}{r} + \frac{j^2}{r^3} - 
\frac{1}{8\pi \rho r^2}\frac{\partial}{\partial r}(r^2 B_{\phi}^2) \nonumber \\ 
& &\mbox{} + \frac{4\eta}{3\rho}\frac{\partial}{\partial r}\left(\frac{1}{r}
\frac{\partial(r\upsilon_r)}{\partial r}\right) \label{vrevolve1} \\
\frac{dj}{dt} & = & \frac{B_r}{4\pi\rho}\frac{\partial}{\partial r}(r B_{\phi})
+\frac{\eta r}{\rho}\frac{\partial}{\partial r}\left(\frac{1}{r}
\frac{\partial j}{\partial r}\right),\label{jevolve1} \\ \label{bphievolve1}
\frac{dB_{\phi}}{dt} & = & -B_{\phi}\frac{\partial\upsilon_r}{\partial r}
+ rB_r \frac{\partial}{\partial r} \left(\frac{j}{r^2}\right), \\ \nonumber 
\frac{d\varepsilon}{dt} & = & -\frac{P}{\rho^2}\left(\frac{d\rho}{dt}\right) 
+\frac{\eta r^2}{\rho}\left(\frac{\partial}{\partial r}\left(\frac{j}{r^2}
\right)\right)^2 \\ \nonumber
&  &\mbox{} + \frac{2\eta}{9\rho}\left[\left(2\frac{\partial \upsilon_r}{\partial r}
-\frac{\upsilon_r}{r}\right)^2 +  r^4 \left(\frac{\partial}{\partial r}
\left(\frac{\upsilon_r}{r^2}\right)\right)^2 \right. \\ & & 
\left. \mbox{} + \frac{1}{r^2}
\left(\frac{\partial(r\upsilon_r)}{\partial r}\right)^2 \right]. 
\label{eintevolve1}
\end{eqnarray}
The terms on the r.\ h.\ s.\ in equation (\ref{eintevolve1}) involving 
derivatives of $\upsilon_r$ and $j$ represent heating due to the 
presence of shear viscosity.  We note that the heating terms which 
depend on $\upsilon_r$ would not be present for an incompressible 
fluid. To handle shocks, we supplement the pressure by an artificial
viscosity term:  $P \to P + q$, where the recipe for $q$ is described
in Appendix~\ref{numerical}.   

\subsection{\em Initial Conditions and Boundary Values}

We assume that no azimuthal component of the magnetic field
is present~initially,
\begin{equation}
B_{\phi}(0,r) = 0, \,\,\,\,  r \geq 0,
\end{equation}
recalling that $B_r$ is always given by equation (\ref{Br}). We are thus
interested in the situation where the azimuthal field is created entirely
by the differential rotation of the fluid, which bends the frozen-in,
initially radial field lines in the azimuthal direction.
Since the region outside of the cylinder is a vacuum, no azimuthal magnetic
field can be carried into this region.  Thus, $B_{\phi}$ must vanish at 
the surface:  
\begin{equation}
B_{\phi}(t,R) = 0, \,\,\,\,  t \geq 0.
\end{equation}
Together with equation (\ref{bphievolve1}), this implies
\begin{equation}
\frac{\partial}{\partial r}\left(\frac{j(t,R)}{r^2}\right) 
= 0, \,\,\,\, t\geq 0.
\label{domegabound}
\end{equation}

The angular velocity $\Omega = j/r^2$ must be finite at the origin.  This
implies that $j(t,0)=0$ for all time.  Also, since the cylinder is 
axisymmetric, the radial velocity at the origin must vanish for all time:
$\upsilon_r(t,0)=0$.  Since we begin our evolutions with an equilibrium
cylindrical polytrope model, we take the radial velocity to be initially
zero everywhere, that is, $\upsilon_r(0,r) = 0$.

We now consider the boundary conditions on the fluid variables at the 
free surface of the cylinder, $r = R$.  Pressure and density must both 
vanish at this~surface:
\begin{eqnarray}
\rho(t,R) & = & 0 \\ \nonumber
P(t,R)   & = & 0, \,\,\,\, t\geq 0.
\end{eqnarray}
When a viscosity is present, the free surface imposes an additional
condition.  Let ${\mathbf n}$ be the unit normal vector to the surface.
Then, at the free surface, balance of forces requires~that
\begin{equation}
-Pn_i + \sigma'_{ij}n_j = 0, \nonumber
\end{equation} 
(Landau \& Lifshitz 1998, eq.\ 15.16).  Since ${\mathbf n}$ is 
radial and $P=0$ on the surface of the cylinder, this condition becomes 
$\sigma'_{rr} = \sigma'_{r\phi}=0$.  Written in component form, 
$\sigma'_{r\phi}=0$ becomes equation (\ref{domegabound}) and 
$\sigma'_{rr}=0$~becomes
\begin{equation}
2\frac{\partial}{\partial r} \upsilon_r(t,R) -\frac{1}{r}\upsilon_r(t,R)
=0, \,\,\,\, t \geq 0.
\end{equation}       

\subsection{\em Conserved Energy and Angular Momentum}

The magnetic Navier-Stokes equation (eq.\ [\ref{navier}]) admits 
two nontrivial integrals of the motion, one expressing conservation 
of energy and the other conservation of angular momentum of the star. 
Energy conservation may be expressed~as
\begin{eqnarray} 
\nonumber
E(t)\ & = & E_{\rm kin}(t)\ +\ E_{\rm mag}(t)\
+\ E_{\rm int}(t)\ +\ E_{\rm grav}(t)\\
& = & E(0),
\label{etot}
\end{eqnarray}
where $E_{\rm kin}(t)$ is the kinetic energy of the matter, 
$E_{\rm mag}(t)$ is the {\em azimuthal} magnetic
energy, $E_{\rm int}(t)$ is the  internal (thermal) energy, 
and $E_{\rm grav}(t)$ is the gravitational potential energy. 
These contributions are as~follows:
\begin{eqnarray}
\label{energies}
E_{\rm kin}(t)  & = & \onehalf \int d\mathcal{A} \rho(t,r)\left(\upsilon_r^2(t,r) +
\frac{j^2(t,r)}{r^2}\right), \nonumber \\
E_{\rm mag}(t)  & = & \int d\mathcal{A} \frac{B^2_{\phi}(t,r)}{8\pi}, \nonumber \\
E_{\rm int}(t)  & = & \int d\mathcal{A} \rho(t,r)\varepsilon(t,r), \nonumber \\
E_{\rm grav}(t) & = & \onehalf \int d\mathcal{A} \rho(t,r)\Phi(t,r), 
\end{eqnarray}
where $\Phi(t,r)$ is the gravitational potential, 
$d\mathcal{A} = 2\pi r dr$, and all energies are taken per 
unit length as in Section~\ref{virsection}.  Equation~(\ref{etot}) 
is derived by rewriting the Eulerian time derivative of the total
energy density using equations (\ref{continuity}), (\ref{navier}), 
and (\ref{einteq}) and applying the divergence theorem. There is no 
explicit contribution from viscous heating in equation~(\ref{etot})
because the internal energy, $E_{\rm int}$, accounts for this. 
We note that $E_{\rm grav}(t)$ is not the same as the gravitational 
term $W$ appearing in the virial theorem for equilibrium cylinders, 
which is defined according to equation (\ref{ints}).  By contrast, 
$E_{\rm grav}(t) = W$ for spherical stars (Shapiro \& Teukolsky 1983).  
Due to the cylindrical geometry of the present problem, $W$ cannot 
be identified as the gravitational potential energy.  However, the 
ratio $T/|W|$, involving quantities from the virial theorem, can 
still be taken as a measure of the degree of rotation for a given 
mass per unit~length.  

Conservation of  angular momentum per unit length is expressed~as
\begin{equation}
J(t)\ = \int d\mathcal{A} \rho(t,r)j(t,r) = J(0)
\label{totangmom}
\end{equation}
Equation (\ref{totangmom}) is obtained by rewriting the Eulerian 
time derivative of the angular momentum density using equations
(\ref{continuity}) and (\ref{navier}) and applying the divergence
theorem.  We note that, consistent with the nonrelativistic MHD 
approximation, the electric field energy $E^2/8\pi$ is not 
included in equation (\ref{etot}) and the angular momentum
of the electromagnetic field ${S}_{\phi}/c^2$, where $\mathbf S$
is the Poynting vector, is not included in 
equation (\ref{totangmom}) (Landau, Lifshitz \& Pitaevskii~2000).

The motivation for monitoring the conservation equations during 
the evolution is twofold: physically, evaluating the individual 
terms enables us to track how the initial rotational energy and 
angular momentum in the fluid  are transformed; 
computationally, monitoring how well the conservation equations 
are satisfied provides a check on the numerical integration~scheme.

\subsection{\em Nondimensional Formulation}

We introduce nondimensional variables to facilitate comparison
of the results of the various dynamical simulations we present in 
Section \ref{results}.  First we define an Alfv\'en velocity using the 
scale of the radial magnetic field from equation~(\ref{Br}):
\begin{equation}
\upsilon_{\rm A} \equiv \frac{B_0}{\sqrt{4\pi\rho_c}},
\end{equation}
where, here and throughout, $\rho_c$ is the {\em initial} central density.
We define nondimensional quantities according~to 
\begin{eqnarray}
\label{nondim}
& & \hat{r} = r/R,\;\;\; \hat{t} = 2t/(R/\upsilon_{\rm A}),\;\;\;
\hat{\eta} = 4\eta/(\rho_c \upsilon_{\rm A} R)
\\ \nonumber
& & \hat{j} = j/\Omega_0 R^2, \;\;\; \hat{B}_\phi =
B_\phi/\left ((4\pi\rho_c)^{1/2}(\Omega_0R) \right) \\ \nonumber
& & \hat{\rho} = \rho/\rho_c, \;\;\;  \hat{\upsilon}_r = 
\upsilon_r/2\upsilon_{\rm A}, \;\;\; \hat{\varepsilon} = 
\varepsilon/(4\pi\upsilon_{\rm A}^2) \\ \nonumber
& & \hat{\mu} = \mu/(2\pi\rho_c R^2),\;\;\; \hat{P} =
P/(8\pi\rho_c\upsilon_{\rm A}^2). 
\end{eqnarray}
Unless stated otherwise, we will work with 
nondimensional variables in all subsequent equations,
but we omit the carets ($\hat{\;}$) on the variables for
simplicity.  With the replacement $V \equiv 1/\rho$  and the 
relation $(\rho r)^{-1}\partial/\partial r = \partial/\partial \mu$,
the system of evolution equations (eqs.\ 
[\ref{densevolve}]-[\ref{eintevolve1}])~becomes
\begin{eqnarray}
\frac{dV}{dt} & = & \frac{\partial}{\partial \mu}(r\upsilon_r) \\
\label{vrevolve2}
\frac{d\upsilon_r}{dt} & = & -2\pi r\frac{\partial P}{\partial \mu}
- \chi\frac{2\mu}{r} -\frac{\zeta_0\chi}{4 r}\frac{\partial}{\partial \mu}
(r^2 B_{\phi}^2) 
\nonumber \\
& & \mbox{} + \left(\frac{\zeta_0\chi}{2}\right)\frac{j^2}{r^3} 
+ \frac{\eta r}{6}\frac{\partial}{\partial \mu}\left(\frac{1}{r}
\frac{\partial(r\upsilon_r)}{\partial r}\right), \\
\frac{dj}{dt} & = & \frac{1}{2}\frac{\partial}{\partial \mu}(r B_{\phi})
+ \frac{\eta r^2}{4}\frac{\partial}{\partial\mu}\left(
\frac{\partial j}{\partial r^2}\right), \\
\frac{dB_{\phi}}{dt} & = & -B_{\phi}\frac{\partial\upsilon_r}{\partial r}
+\frac{1}{2}\frac{\partial}{\partial r}\left(\frac{j}{r^2}\right), \\
\frac{d\varepsilon}{dt} & = & -\frac{P}{\rho^2}\frac{d\rho}{dt} 
 + \frac{\zeta_0\chi}{16\pi}\frac{\eta r^2}{\rho}\left(
\frac{\partial}{\partial r}\left(\frac{j}{r^2}\right)\right)^2 \nonumber \\
& & \mbox{} + \frac{1}{8\pi}\frac{2\eta}{9\rho}\left[
\left(2\frac{\partial \upsilon_r}{\partial r}
-\frac{\upsilon_r}{r}\right)^2 + r^4 \left(\frac{\partial}{\partial r}
\left(\frac{\upsilon_r}{r^2}\right)\right)^2 \right. \nonumber \\
& & \mbox{} + \left.\frac{1}{r^2}
\left(\frac{\partial(r\upsilon_r)}{\partial r}\right)^2 \right].
\label{eintevolve2}
\end{eqnarray}  
Here, $\zeta_0 = \zeta(r=0)$ is as defined in equation 
(\ref{zetadef}), $\chi \equiv \Omega_0^2R^2/2\upsilon_{\rm A}^2$, 
and artificial viscosity is incorporated in $P$ as described in 
Appendix~\ref{numerical}.

\section{DYNAMICAL EVOLUTION}
\label{results}

The MHD evolution cases which we will discuss in Sections
\ref{incomp}--\ref{n5results} are
specified by four parameters: the polytropic index $n$, the 
rotation parameter $\zeta_0 = \Omega_0^2/\pi\rho_c$, the Alfv\'en
speed $\upsilon_{\rm A} = B_0/\sqrt{4\pi\rho_c}$, and the 
coefficient of dynamic viscosity
$\eta$.  We will give results for $n = 0.001,\,1,\,3,\,$ and 5 in
order to probe the effects of the degree of central condensation.
For $n=0.001$ and $n=1$, all differentially rotating cylinders
have $\beta$ smaller than the maximum value allowed for uniform
rotation, $\beta_{\rm max}$.  For these cases, 
$\zeta_0$ is chosen so that a moderate
value of $\beta = T/|W|$ is obtained.  For differentially 
rotating cylinders with $n=3$ and $n=5$ however, it is
possible to choose values of $\zeta_0$ for which 
$\beta > \beta_{\rm max}$. 
Thus, for both of these cases, we give results for two 
values of $\beta$, one with $\beta > \beta_{\rm max}$ 
and one with $\beta < \beta_{\rm max}$.  Realistic cold neutron
star EOS's are expected to be fairly stiff, corresponding to 
polytropes with $0.5 \lesssim n \lesssim 1.0$.  For polytropic
spheres, this range of $n$ corresponds to 
$1.84 \leq \rho_c/\bar{\rho} \leq 3.29$.  To obtain the same
range of $\rho_c/\bar{\rho}$ for cylinders requires
$0.70 \leq n \leq 1.49$.  Our $n=1$ model lies in this range,
but we will also study the more compressible models since
these allow $\beta > \beta_{\rm max}$ with our adopted angular 
velocity profile and since we wish to
understand the effects of compressibility.  We also note
that our results may be relevant for magnetic braking in
radiation pressure dominated SMS's, which have $n \approx 3$
and $\rho_c/\bar{\rho} \approx 54$.      

For each case below, $\upsilon_{\rm A}$ is chosen so that 
$\mathcal{M}_{\rm A}/T = 7.85 \times 10^{-3}$, where 
$\mathcal{M}_{\rm A}$ gives the relative scale of the energy 
per unit length associated with the initial radial 
magnetic~field:
\begin{equation}
\mathcal{M}_{\rm A} \equiv \pi R^2 \frac{B_0^2}{8\pi} =
\onehalf \pi\rho_c R^2 \upsilon_{\rm A}^2.
\label{Mdefn}
\end{equation}
Because of the static line singularity at $r=0$, the actual
energy associated with the radial field $B_r$ is not well 
defined.  For this reason, the energy scale 
$\mathcal{M}_{\rm A}$ is used to characterize the radial 
field strength.  We choose the ratio $\mathcal{M}_{\rm A}/T$ 
to be the same for all of our runs because this ratio roughly 
determines the extent to which the radial field lines are 
wound up.  Thus, keeping this ratio constant facilitates 
comparison between the different evolution cases.  Choosing 
$\mathcal{M}_{\rm A} \ll 1$ shows the consequences of an 
initially {\em weak} magnetic field in influencing the dynamical 
behavior of an equilibrium star.  This is the situation of 
greatest astrophysical interest for core collapse, 
hypermassive neutron star evolution, and other relativistic MHD
scenarios.  We define an Alfv\'en wave crossing time as 
$t_{\rm A} = R/\upsilon_{\rm A}$, which in nondimensional units 
is just $t_{\rm A} = 2$.  Thus, the basic time unit for the 
evolution code is the Alfv\'en~time.

Lastly, for cases with shear viscosity, the coefficient of dynamical
viscosity is chosen as $\eta = 0.2$ in nondimensional units.  This 
intermediate value is chosen in order that the viscosity in the star
be sufficiently large for dissipation to become evident in a 
few Alfv\'en timescales, but sufficiently small so that viscous 
damping of differential rotation does not completely suppress
the growth of the azimuthal magnetic field.  The characteristic 
viscous dissipation timescale can be taken as $t_{\eta} = 1/\eta$
in our nondimensional units, which is equivalent to $t_{\eta} = 
R^2\rho_c/8\eta$ in physical units (see eq.\ [\ref{nondim}]). 
The choice $\eta = 0.2$ corresponds to the ratio of the 
viscous to Alfv\'en timescales $t_{\eta}/t_{\rm A} = 2.5$.  

To illustrate the hierarchy of timescales governing the evolution 
of relativistic rotating stars, we will evaluate the relevant 
timescales for neutron star parameters consistent with the stable 
hypermassive remnant found in the simulations of Shibata \& 
Ury\=u (2000).  The dynamical timescale associated with the remnant 
is given~by
\begin{equation}
t_{\rm dyn} = \frac{1}{\sqrt{\rho_c}}
\approx 0.17 \left({R \over 20 \mbox{km}}\right)^{3/2} 
\left({M \over 3 M_{\odot}}\right)^{-1/2} \mbox{ms}, 
\label{tdyn}
\end{equation}
the time it takes the binary remnant to achieve equilibrium 
following coalescence.
This is also the time that it takes the star, if it is 
driven far out--of--equilibrium 
by magnetic braking, to undergo collapse. The central rotation period 
of the remnant~is
\begin{equation}
t_{\rm rot} = {2\pi \over \Omega_0} 
\approx 0.3 \left({R \over 20 \mbox{km}}\right)^{3/2} 
\left({M \over 3 M_{\odot}}\right)^{-1/2} \mbox{ms}, 
\label{for}
\end{equation}
while the period at the equator is about three times longer. The 
timescale for magnetic braking of differential rotation by
Alfv\'{e}n waves is given~by  
\begin{equation}
t_ {\rm A} = {R \over v_{\rm A}}\approx 10^2 \left({B_0 \over 10^{12} 
  \mbox{G}}\right)^{-1}
\left({R \over 20 \mbox{km}}\right)^{-1/2}
\left({M \over 3 M_{\odot}}\right)^{1/2} \mbox{s}.
\label{forone}
\end{equation}
On this timescale, the angular velocity profile in the star is
significantly altered. Finally, viscous dissipation drives the 
star to a new, uniformly rotating equilibrium state on a~timescale
\begin{equation}
t_{\eta} \approx 2\times 10^9 \left({R \over 
20 \mbox{km}}\right)^{23/4}\left({T \over 10^9 {\mbox K}}\right)^2
\left({M \over 3 M_{\odot}}\right)^{-5/4} \mbox{s}  , 
\label{fortwo}
\end{equation}
where $\eta = 347 \rho^{13/4} T^{-2} {\rm cm^2 s^{-1}}$ (Cutler \& 
Lindblom 1987).  In our simulations we choose parameters 
to preserve the inequality $t_{\rm dyn} < t_{\rm rot} < t_{\rm A} 
< t_{\rm \eta}$, although the relative magnitudes are altered 
for numerical~tractability.

In section \ref{incomp}, we will discuss results of dynamical
simulations with $n = 0.001$ and compare with the results of 
Paper I.  Then in section \ref{n3results}, we treat $n=3$ 
cylinders in detail as these cases display typical compressible
behaviors.  Sections \ref{n1results} and \ref{n5results} will 
briefly review results of dynamical simulations for $n=1$ and 
$n=5$, respectively.  Appendix \ref{numerical} summarizes our 
numerical method and difference~equations. 
 
\subsection{\em Nearly Incompressible Cylinders}
\label{incomp}
As a check of our code, we first reproduce the analytic results 
of Paper I for incompressible cylinders by considering a small 
value of the polytropic index: $n=0.001$, which corresponds to 
$\Gamma = 1001.0$.  This extremely stiff equation
of state is a good approximation to an incompressible fluid.
The polytropic index $n$ was not simply taken to zero because 
$n=0$ corresponds to infinite sound speed, $c_s$.  The timestep 
in our simulations is limited by the Courant stability criterion 
$\delta t < \delta r/c_{\rm s}$, where $\delta r$ is the 
smallest spatial grid interval. Thus our numerical method is 
not applicable to the strictly incompressible $n=0$~case. 

\begin{figure}
\plotone{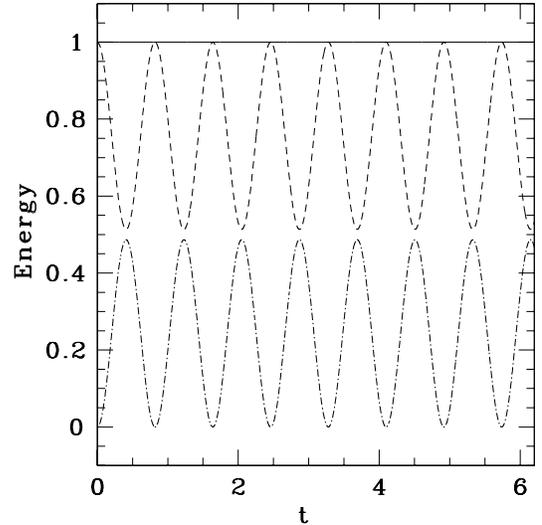}
\caption{Energy evolution for a differentially rotating, nearly
  incompressible star with zero viscosity and a vacuum exterior 
  (Case~I).  The dashed line shows $E_{\rm kin}$ and the dot-dash 
  line shows $E_{\rm mag}$ ($E_{\rm grav}$ and $E_{\rm int}$ are 
  essentially constant and are not shown).  All energies are 
  normalized to $E_{\rm kin}(0)$.  The sum of $E_{\rm kin}$ and 
  $E_{\rm mag}$ is conserved and remains equal to the initial 
  rotational energy $E_{\rm kin}(0)$, which is plotted as the 
  solid horizontal line at Energy = 1.  Time is in nondimensional 
  units according to equation~(\ref{nondim}).\label{balIeta0}}
\end{figure}

\begin{figure}
\plotone{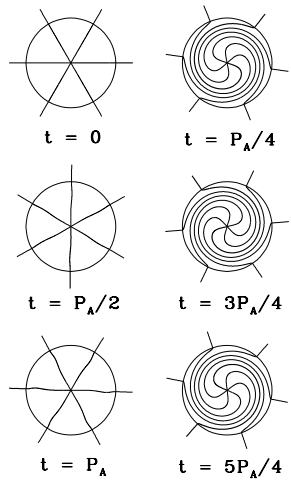}
\caption{Magnetic field line configurations for Case I (nearly 
  incompressible, $n=0.001$, zero viscosity) at selected times.  The 
  twisting of the field lines is due to differential rotation.  
  Snapshots are given at time intervals of $P_{\rm A}/4$, where 
  $P_{\rm A}$ is the standing Alfv\'en wave period (see 
  eq.\ [\ref{PA}]). Each frozen-in field line passes through the 
  same fluid elements at all times.  The bold field line is an 
  arbitrarily chosen fiducial~line. \label{twistIeta0}}
\end{figure}

Results for Case I ($n=0.001$) with zero shear viscosity 
($\eta=0.0$) are shown in Figs.\ \ref{balIeta0} and
\ref{twistIeta0}.  In Fig.\ \ref{balIeta0}, we track the evolution 
of various energies with time and observe how the contributions
to the total conserved energy oscillate between rotational kinetic 
energy of the fluid and the energy of the azimuthal magnetic field.  
The angular momentum is also strictly conserved.  The period of 
these oscillations, $P_{\rm A} = 0.82\, t_{\rm A} = 1.64$ also agrees 
with the analytical result.  We will henceforth~define
\begin{equation}
P_{\rm A} \equiv 0.82\, t_{\rm A} \label{PA}
\end{equation}
so that the actual value of the timescale $P_{\rm A}$ will
depend on the parameters of the model in question.  
The dynamical time in this case, $t_{\rm dyn} = 0.13$, is much 
shorter than the Alfv\'en wave oscillation period.  Thus, the 
conversion of rotational to magnetic energy occurs over several
dynamical timescales. We note that the outer radius of the cylinder 
remains constant to around one part in $10^4$, showing that
the effects of the slight degree of compressibility are not important
for this simulation.  There is essentially no radial motion and the 
cylinder remains very close to virial equilibrium.  Let the 
normalized virial sum be defined~as 
\begin{equation}
\mathcal{V} \equiv \frac{2\Pi + 2T + W}{|W|},
\label{virsum}
\end{equation}  
since $|W|$ is constant (see eqs.\ [\ref{vir}] and [\ref{Wint}]).  For 
this evolution, $\mathcal{V} \lesssim 1.5 \times 10^{-5}$ 
($\mathcal{V} = 0$ for equilibrium configurations).  As an additional 
check, we found that the rotational and magnetic energies at times 
$t = P_{\rm A}/4, 3P_{\rm A}/4, \dots$ when the azimuthal magnetic 
field reaches its maximum magnitude, are as follows: 
$E_{\rm kin}/E_{\rm kin}(0) = 0.5136$ and 
$E_{\rm mag}/E_{\rm kin}(0) = 0.4865$.  These agree well 
with the analytical results given in Paper I of 0.513 and 
0.487, respectively.  


A crucial property of the analytic solution for the 
case $n=0$ is the scaling behavior (see Paper I).  For that solution, the 
{\em amplitudes of all evolved quantities are entirely independent
of the initial radial seed field} given in equation (\ref{Br}).
{\em Only the timescale of the oscillations of the energy 
components depends on $B_0$} and this timescale is proportional
to $B_0^{-1}$.
We confirmed that our numerical results for $n=0.001$ also 
satisfy this scaling behavior to high accuracy. This check 
was performed by doubling $B_0$ and observing that the oscillation
period halves: $P_{\rm A} \to P_{\rm A}/2$.  This scaling is
not expected to hold for general values of $n$, however, because
input parameters characterizing the initial rotation and magnetic
field enter explicitly in the evolution system 
(see eqs.\ [\ref{vrevolve2}]-[\ref{eintevolve2}]).  However, the
analytic scaling should be approximately valid for stiff equations 
of state in a first~approximation.

The tradeoff between rotational kinetic and magnetic energies, 
$E_{\rm kin}$ and $E_{\rm mag}$, seen in Fig.\ \ref{balIeta0} matches
the standing Alfv\'en wave behavior seen in Paper I.  
First the differential rotation
generates a nonzero $B_{\phi}$ which drains energy away from 
the differential rotation.  When $B_{\phi}$ reaches its maximum, the
cylinder is uniformly rotating, and the built-up magnetic stress 
begins to drive differential rotation in the opposite direction, 
unwinding the  magnetic field and then winding it up again in 
the opposite sense.  This cycle corresponds to a standing 
Alfv\'en wave with period $P_A$. 
This process is shown explicitly in Fig.\ \ref{twistIeta0}, which gives 
cross-sectional views of representative magnetic field lines
at critical phases during an oscillation period.  The field lines were
created by integrating the equation $r d\phi/dr = B_{\phi}/B_r$ to find 
$\phi = \phi(r)$.  (Our 1+1 Lagrangian evolution determines
$B_{\phi}$ and the Eulerian coordinate $r$ for a given fluid
element, while $B_r$ is fixed.)  
At times $P_{\rm A}/4$ and $3P_{\rm A}/4$ corresponding 
to the maximum magnitude 
of $B_{\phi}$, the field lines are highly twisted.  The degree to which
the field lines are twisted depends on the magnitude of the initial 
radial magnetic field (see eq.\ [\ref{Mdefn}] and associated 
discussion).  Even for a small initial field, however, the 
azimuthal magnetic field will grow to the {\em same} high value sufficient 
to brake the differential motion and drive the star to oscillate 
about the state of uniform rotation. This will have important 
consequences for a hypermassive star which depends on differential 
rotation for stability against gravitational~collapse.


\begin{figure}
\plotone{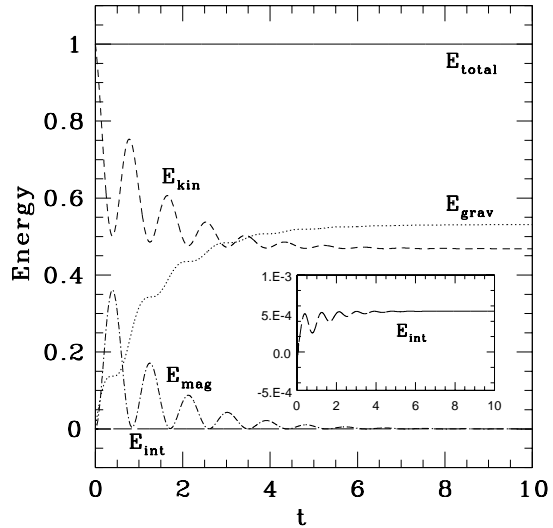}
\caption{Energy evolution for Case I ($n = 0.001$)  
  with shear viscosity given by $\eta = 0.2$.  The dotted line 
  labeled ``$E_{\rm grav}$'' is actually 
  $(E_{\rm grav}-E_{\rm grav}(0))/E_{\rm kin}(0)$.  
  The line labeled ``$E_{\rm int}$'' is also defined in this way.  
  The expansion of the cylinder causes the gravitational 
  potential energy to become less negative, compensating for the loss 
  of $E_{\rm kin}$.  The sum of the energies is conserved and is plotted 
  as the horizontal solid line labeled~``$E_{\rm total}$.'' 
  \label{balIeta}}
\end{figure}

\begin{figure}
\plotone{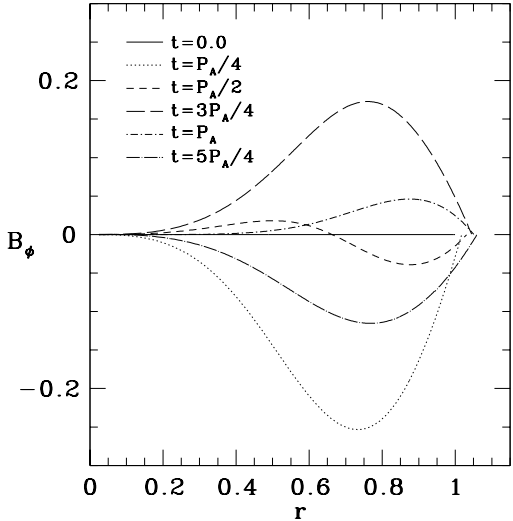}
\figcaption{Azimuthal magnetic field profiles for Case I with viscosity.
  The radius $r$ and azimuthal field $B_{\phi}$ are plotted in nondimensional
  units according to equation (\ref{nondim}).  Snapshots of the profiles are
  taken at the same times as in Fig.\ \ref{twistIeta0}.  The behavior is 
  approximately periodic, with a period similar to $P_{\rm A}$.  Because 
  of the viscosity, oscillations of the $B_{\phi}$ profile are damped.
  Note that the profiles do not meet at the radius of the outer mass shell
  due to the slight expansion of the~cylinder.  \label{bprofIeta}}
\end{figure}

Next, we consider an approximately incompressible model with nonzero
shear viscosity (Case I with $\eta=0.2$), and results are given in
Figs.\ \ref{balIeta} and \ref{bprofIeta}.  In Fig.\ \ref{balIeta},
we track the 
evolution of the various energies with time.  We observe how the 
oscillations of the rotational kinetic and azimuthal magnetic 
field energies are now damped by viscosity.  The damping takes 
place over a few periods ($P_A$) of the magnetic field oscillations, 
consistent with the choice of $\eta=0.2$ and hence 
$t_{\eta}/t_A = 2.5$.  This damping behavior was also seen in  
the results of Paper I.  However, in this case, the heating due to viscosity 
causes the cylinder to expand, with the outer radius expanding by 
about 7\%.  The expansion causes a decrease in the magnitude of the 
(negative) gravitational potential energy $E_{\rm grav}$ which 
compensates for the loss of rotational kinetic energy due to the viscous 
dissipation.  (We note that the line labeled ``$E_{\rm grav}$'' in 
Fig.\ \ref{balIeta} is actually $(E_{\rm grav}-E_{\rm grav}(0))/E_{\rm kin}(0)$.) 
In contrast, the internal energy does not change much from its initial 
value.  We note that total energy and angular momentum are conserved 
very well for this simulation (to one part in $10^7$ and $10^9$, 
respectively).  The fact that the cylinder expands in this case clearly
indicates that, when there is a significant amount of viscous heating, 
compressibility effects are significant even for $n=0.001$.  
To confirm that this expansion is indeed caused by heating, we performed
this simulation a second time without the viscous heating terms ({\em i.e.}
by removing the terms proportional to $\eta$ in eq. [\ref{eintevolve2}]).  
No expansion occurs in this case.  When viscous heating is included,
the pressure term $\Pi$ in the virial equilibrium equation (eq. [\ref{vir}])
significantly increases. This change is not reflected in the internal energy, 
however, since $E_{\rm int} = n \Pi$ (by eqs. [\ref{ints}] and 
[\ref{energies}] and the $\Gamma$-law EOS) and $n$ is small in this~case. 


\subsection{\em The ${\mathbf n=3}$ Results}
\label{n3results}
\begin{figure}
\plotone{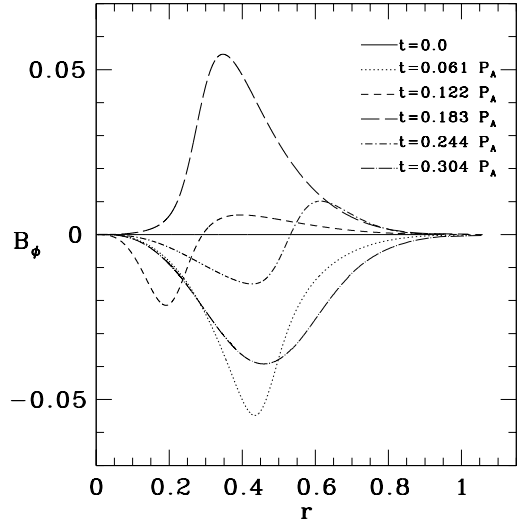}
\figcaption{Azimuthal magnetic field profiles for Case II ($n~=~3$ with slow
  differential rotation) for $\eta = 0$.  The radius $r$ and azimuthal 
  field $B_{\phi}$ are plotted in nondimensional units according to equation 
  (\ref{nondim}).  Snapshots of the profiles are taken at steps of $0.061\, P_A$.  
  The behavior is seen to be  quasi-periodic, with a period of 
  $\sim 0.244\, P_A$.  For example, the maximum magnitudes of $B_{\phi}$ 
  occur approximately at fractions $1/4, 3/4,$ and $5/4$ of the period. 
  This is the same period seen in the gentle radial oscillations which 
  occur during the evolution of this~model. \label{bprofIIeta0}}
\end{figure}

We now present results for the slowly differentially rotating $n=3$ case
(Case II) without viscosity, and with $\beta = 0.030$ ($< \beta_{\rm max}$). 
The numerical evolution for this model shows a slight overall expansion of 
the cylinder and radial oscillations which increase in magnitude toward 
the surface. Oscillations with the same period are also seen in the radial 
profile of $B_{\phi}$, as shown in Fig.~\ref{bprofIIeta0}, which shows 
snapshots of the $B_{\phi}$ profile at selected phases of the 
quasi-oscillation period.  This period is $\sim 0.244\, P_{\rm A}$, 
significantly shorter than the Alfv\'en period, $P_{\rm A}$, as defined in 
equation~(\ref{PA}).  The dynamical timescale for this case, 
$t_{\rm dyn} = 0.0091$ is much shorter than the period of the radial 
oscillations (since $0.244\,P_{\rm A} = 0.400$), and the oscillatory behavior 
is therefore quasistatic. The angular velocity profiles for this model also 
display this oscillatory behavior, with the angular velocity becoming more 
uniform at phases corresponding to maxima in $B_{\phi}$.  The radial 
oscillations show no indication of decaying during the simulation, which 
ran for several Alfv\'en timescales: 
$t_{\rm final} \sim 3\, t_{\rm A} = 3.7\, P_{\rm A}$. Thus, with no dissipation 
mechanism, these oscillations presumably continue indefinitely.  [Note, 
however, that a small amount of viscosity was added in order to stabilize 
the run, so that $t_{\eta}=70.0\, t_{\rm A}$.  This has no significant impact  
on this run, since the total length of the run was only $\sim 3\, t_{\rm A}$.] 
That this model does not display any strong radial motions is expected 
since $\beta = T/|W|$ is below the limit for uniform rotation and hence 
the configuration can be driven to uniform rotation quasi--periodically 
without undergoing appreciable contraction and/or expansion. 


Now suppose a significant shear viscosity is introduced to the case 
described in the previous paragraph.  As in Case II without viscosity, the 
cylinder expands slightly.  Radial oscillations occur again with roughly 
the same period ($P \approx 0.244\, P_A$), but these oscillations are damped 
after a few periods due to the shear viscosity.  The final state is 
uniformly rotating with no azimuthal magnetic field.  In this case, the 
cylinder easily relaxes to uniform rotation and dynamical equilibrium 
within a few oscillation periods of the magnetic field.  The excess 
rotational energy is converted into internal energy~(heat).

\begin{figure*}
\plottwo{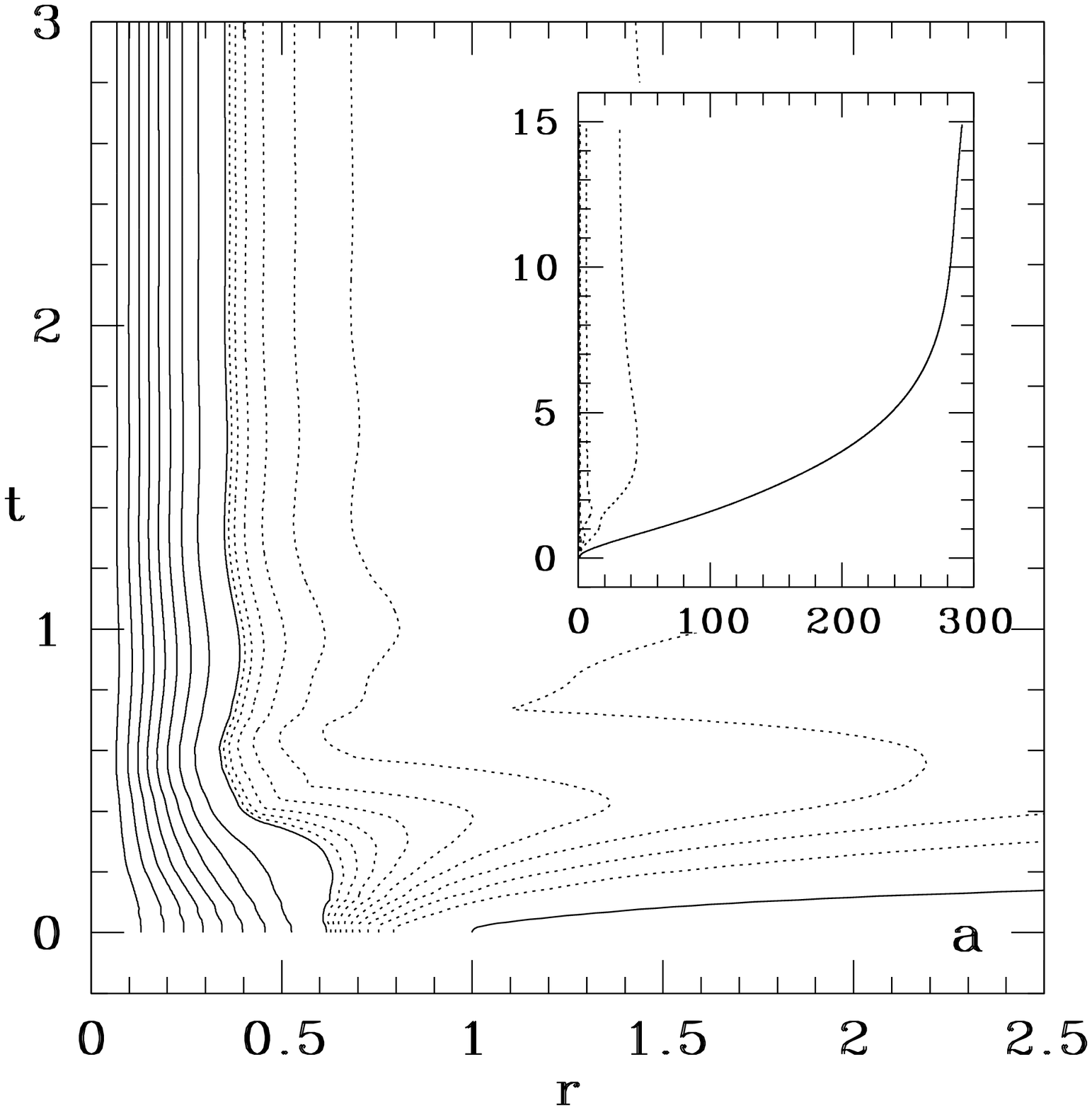}{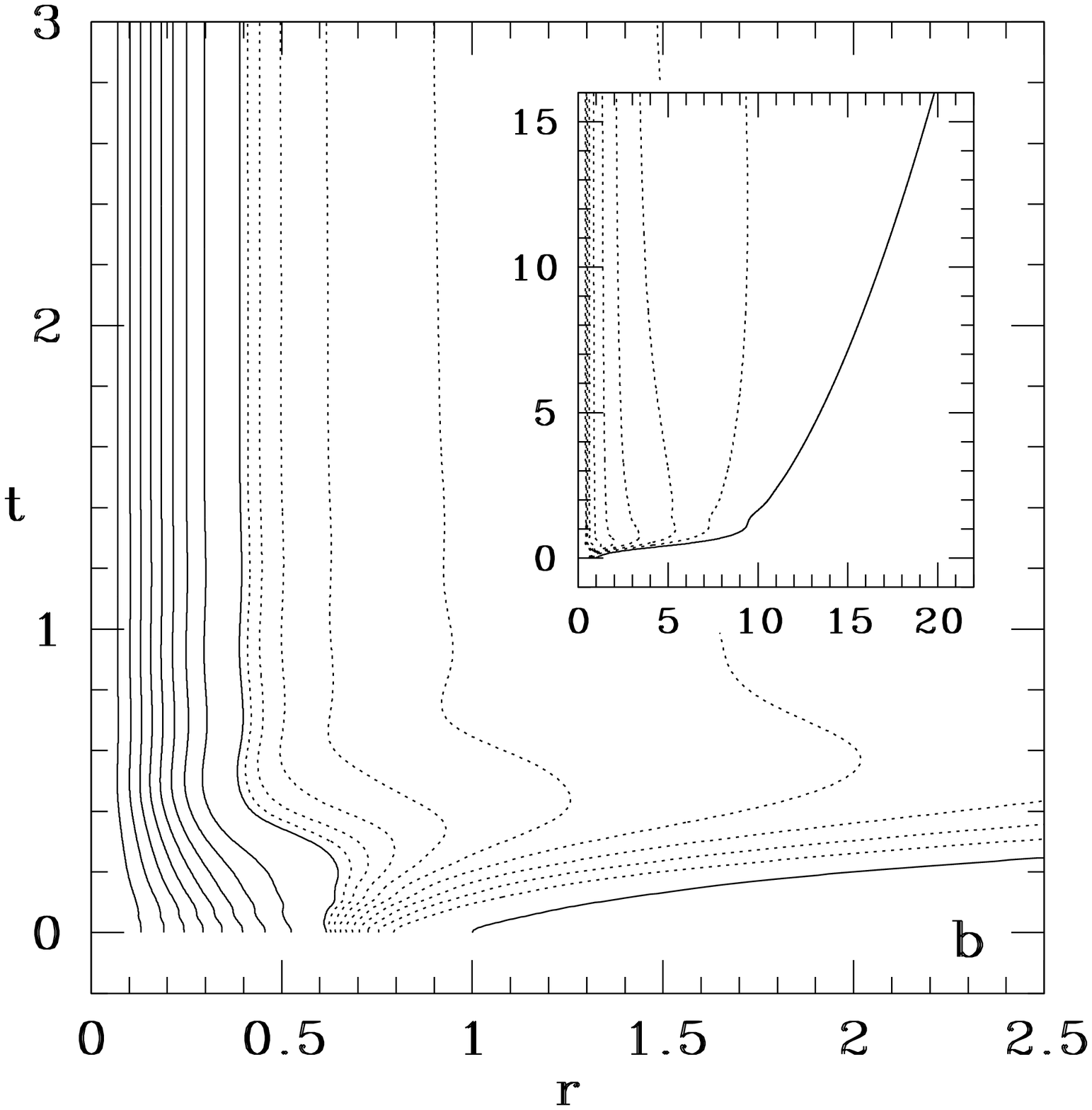}
\figcaption{(a) Spacetime diagram featuring Lagrangian mass 
  tracers for Case III ($n=3$, rapid differential rotation) with $\eta = 0$.  
  The solid lines show mass fractions $0.10,0.20,\dots,1.0$ while the dotted 
  lines enclose fractions $0.91, 0.92,\dots 0.99$.  The large plot 
  demonstrates the contraction of the inner shells and the expansion of the 
  outer envelope.  The inset shows the same outermost tracers out to a much 
  larger radius and for a longer time.  This inset shows that a diffuse 
  atmosphere has been formed outside of the cylinder.
  (b) Same as (a) except now with viscosity $\eta = 0.2$. The 
  contraction of the inner shells and expansion of the outer envelope shown 
  here are milder than the corresponding behaviors in (a) with 
  zero~viscosity. \label{tracerIII}}
\end{figure*}

We now examine the results for the rapidly differentially rotating 
$n=3$ case (Case III) with $\eta = 0$ and  
$\beta = 0.191 \gg \beta_{\rm max}$.  Though we are interested in 
the behavior with no shear viscosity, we again add a very small shear 
viscosity to stabilize the simulation.  In this case, taking $\eta$ such 
that $t_{\eta}=500.0\, t_{\rm A}$ is sufficient, and the effects of the 
shear viscosity will not be seen in the following results.  In all cases
discussed hereafter, we will add a similar small, stabilizing viscosity
that does not affect the results on the timescales of interest.  (This
small viscosity will, of course, not be added in cases where we are
considering the effects of a significant shear viscosity with $\eta$ 
chosen such that $t_{\eta} = 2.5\, t_{\rm A}$.) The motion of 
the Lagrangian mass tracers for this case is shown in Fig.~\ref{tracerIII}(a) 
(Fig.~\ref{tracerIII}(b) shows the analogous results with significant
shear viscosity and will be discussed below). The inner $\sim 95\% $ of 
the mass undergoes a significant contraction and a slight bounce. The 
inset figure shows that the outer $\sim 5\% $ of the mass expands to 
large radii, forming a diffuse atmosphere.  In the final state, the 
interior of the star is slowly rotating, having lost most of its 
angular momentum to the outer~envelope.  


\begin{figure}
\plotone{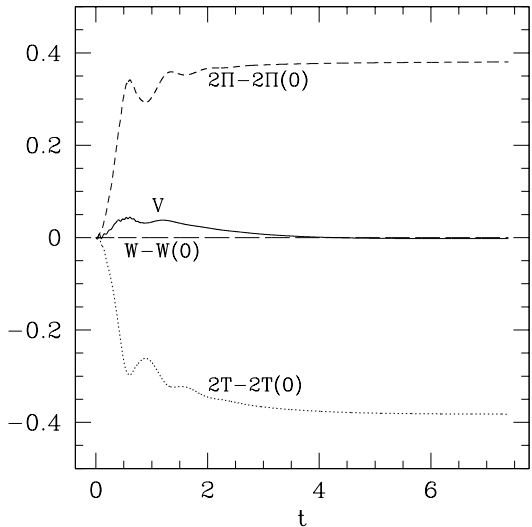}
\caption{Individual components of the virial sum $\mathcal{V} = 
  (2\Pi + 2T + W)/|W|$ for Case III ($\mathcal{V}=0$ in strict equilibrium)
  with $\eta = 0$.  The components $2\Pi$, $2T$, and $W$ are normalized 
  to $|W|$ and are plotted with their initial values subtracted off. That 
  $\mathcal{V}$ remains small while the cylinder undergoes significant 
  contraction indicates that the contraction and readjustment occur in 
  a quasistatic~fashion. \label{virialIII}}
\end{figure}

Figure~\ref{tracerIII}(a) shows that the duration of the contraction is 
approximately the Alfv\'en timescale, $t_{\rm A} = 2$.  The dynamical 
timescale as defined in equation (\ref{tdyn}) is much shorter 
than the Alfv\'en timescale:
$t_{\rm A} = 66.4\, t_{\rm dyn}$.  Since the contraction takes place over
many dynamical timescales, the cylinder is never far from equilibrium
and the evolution is quasistatic.  The evolution of the virial sum 
(eq.\ [\ref{virsum}]), which is shown in Fig.\ \ref{virialIII}, illustrates
this fact.  While the pressure and kinetic energy terms of the virial sum 
depart significantly from their initial values, the sum $\mathcal{V}$ 
remains near its equilibrium value of zero. Also, typical radial velocities 
of the contracting shells have magnitudes only $\sim 10 \%$ of the 
``free-fall velocity,'' $\upsilon_{\rm ff}$, where 
$\upsilon_{\rm ff} \equiv R/t_{\rm dyn}$.  In a realistic relativistic 
hypermassive star, this quasistatic contraction would likely lead to 
dynamical radial instability and then catastrophic gravitational collapse 
of the core. The diffuse atmosphere seen in our simulation suggests that 
magnetic braking operating in rotating stars may also lead to the ejection 
of winds or a diffuse ambient~disk.  

\begin{figure}
\plotone{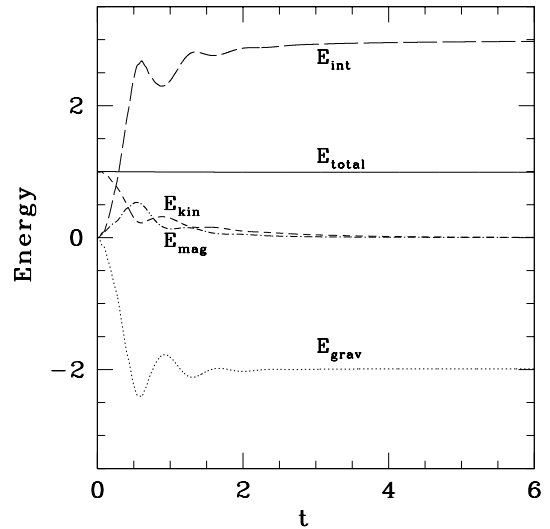}
\figcaption{Energy evolution for Case III ($n=3$ and rapid differential 
  rotation) with $\eta = 0$.  All energies are defined as in Fig.\ \ref{balIeta}.  
  The rapid rise in $E_{\rm int}$ and sharp drop in $E_{\rm grav}$ 
  correspond to the contraction of the core.  The magnetic energy 
  $E_{\rm mag}$ grows to an appreciable fraction of the initial 
  kinetic energy before $B_{\phi}$ decays back to zero in the 
  final~state. \label{balIIIeta0}}
\end{figure}

\begin{figure}
\plotone{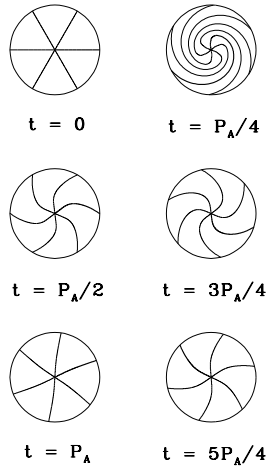}
\caption{Magnetic field line configurations for Case III ($n=3$, 
  rapid differential rotation) with $\eta = 0$ as in Fig.\ \ref{twistIeta0}.  
  The magnetic field configuration at each time is shown only for the inner $96\%$ 
  of the mass because the outer envelope is ejected to large radius, forming
  a low-density atmosphere (hence the continuation of each field line 
  outside the cylinder is not shown).  The azimuthal magnetic field is clearly 
  strongest at early~times. \label{twistIIIeta0}}
\end{figure}

The time evolution of the various components of the energy is shown in 
Fig.\ \ref{balIIIeta0}.  The most salient feature is the strong contraction, 
which results in the sharp increase in $E_{\rm int}$ and an increase in the 
{\em magnitude} of the gravitational potential energy.  One also sees that, 
early in the simulation, the magnetic energy grows to an appreciable 
fraction of the initial rotational kinetic energy.  In the final state, 
most of the kinetic energy has been converted into internal energy by the 
contraction.  Angular momentum is nonetheless conserved because the outer 
shells are very slowly rotating at very large radii. We show cross-sectional 
views of the magnetic field lines in Fig.\ \ref{twistIIIeta0}.  This shows
that the maximum magnitude of $B_{\phi}$ occurs early in the run, with some 
small oscillations following.  There is a striking contrast in behavior 
between this case with rapid differential rotation and Case II discussed 
above.  For Case II, $\beta < \beta_{\rm max}$ and the magnetic field 
(and viscosity, when present) can drive the cylinder toward uniform rotation 
without significantly altering the structure of the cylinder.  In Case III, 
however, $\beta$ is significantly larger than $\beta_{\rm max}$ and the 
cylinder cannot relax to uniform rotation without restructuring.  In this 
case, we see both a strong contraction and expansion of the outer layers.  
Thus, when the differential rotation is strong, magnetic braking leads to 
dramatic changes in the stellar~structure.


\begin{figure}
\plotone{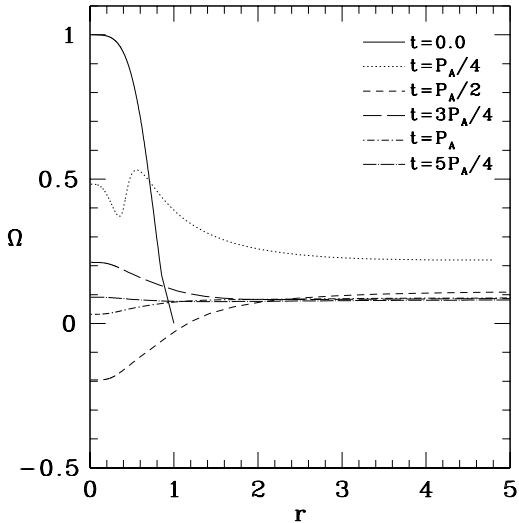}
\caption{Angular velocity profiles for Case III with 
  $\eta = 0.2$ at selected times.  The configuration is driven
  toward slow, uniform rotation.  Later profiles extend to $r>1$ 
  because of the expansion of the cylinder. Angular velocity and 
  time are expressed in nondimensional units according to 
  equation~(\ref{nondim}).  Though the final uniform value of 
  $\Omega$ is much smaller than the initial values in the 
  interior, angular momentum is conserved due to the slow 
  rotation of the outer shells at large~radii. 
  \label{omprofIIIeta}}
\end{figure}

These results can be compared with the analogous case with significant
shear viscosity present ($\eta = 0.2$).  Figure~\ref{tracerIII}(b) shows 
the Lagrangian matter tracers for this case.  A contraction similar to that 
seen in Case III with $\eta = 0$ is seen again here. However, the inset 
plot shows that the outer shells are not ejected to such large distances 
as they are in the previous case.  Also, a strong bounce is not seen in 
the present case.  One may conclude that the presence of viscosity has 
resulted in slightly milder behavior.  This is also seen in the fact that
the central density increases by a factor of 3.50 in this case, but 3.71 in 
the case without significant shear viscosity. The approach to uniform 
rotation in this case is easily seen in Fig.\ \ref{omprofIIIeta}, which 
gives angular velocity profiles at various times.  As before, the final 
state has very little rotational kinetic energy and angular momentum is 
conserved by transferring angular momentum (via viscosity and magnetic 
fields in this case) to slowly rotating outer shells at large~radii.  

The final state for Case III with $\eta = 0$ has 
$\langle P/\rho^{\Gamma} \rangle/\langle P/\rho^{\Gamma} \rangle_{t=0} = 1.11$, 
which indicates that the 
strong contraction and bounce resulted in significant shock heating.  
[The symbol $\langle P/\rho^{\Gamma} \rangle$ represents an average of 
$P/\rho^{\Gamma}$ over the inner 96\% of the mass.  The average was
restricted to the inner 96\% because $P/\rho^{\Gamma}$ could not be 
calculated accurately in the low-density atmosphere.]  
The existence of shocks is not in conflict with the overall
quasistatic nature of the evolution because all of the shocks that we
have observed are weak (with Mach numbers in the range $1.0 <
\mathcal{M} \lesssim 2$) and occur in the outer, low density regions.
Consider the 
shock which formed during this evolution at $t \sim P_{\rm A}/4$ at 
$\mu(r)/\mu_t \sim 0.93$.  The MHD shock propagated outward into the 
atmosphere of the cylinder, and the incoming fluid had a slightly 
supersonic inward radial velocity with respect to the shock front. 
The angular velocity and azimuthal magnetic 
field had discontinuously larger values at the front in the unshocked 
fluid versus the shocked fluid.  This is because the shells falling into 
the shock front spin up due to momentum conservation.  The resulting 
discontinuity in $\Omega$ then leads to a discontinuity in $B_{\phi}$ 
due to flux freezing.  
(In Section~\ref{n5results}, we will give an example of a shock for which
$B_{\phi}$ actually changes sign across the shock front.)
We explored the extent to which the quantities 
upwind and downwind of this shock satisfy the jump conditions for shocks 
in magnetic fluids (Landau, Lifshitz, \& Pitaevskii 1984). For example, 
the following condition can be derived from continuity of the 
$z-$~component of the electric field (which applies for a steady state 
shock) and conservation of mass across the MHD shock~front:
\begin{equation}
\rho_1 \left( \frac{B_{\phi,2}}{\rho_2}-\frac{B_{\phi,1}}{\rho_1}
\right) = \frac{B_r}{\upsilon_{r,1}}(\upsilon_{\phi,2}-
\upsilon_{\phi,1}),
\label{shock}
\end{equation}
where the subscripts ``1'' and ``2'' refer to the pre- and post-shock
fluids respectively (this equation is in physical units, not 
nondimensional units).  The difference between the right and left hand
sides of equation (\ref{shock}) for the shock in our numerical 
solution is $ \sim 8\%$.  The other jump conditions are also satisfied 
with errors of $\lesssim 20 \%$.  This agreement is reasonable 
considering the simplicity of the artificial viscosity scheme employed 
by our code and the fact that the jump conditions with which we 
compared our results assume steady state rather than 
dynamical conditions.

\subsection{\em The ${\mathbf n=1}$ Results}
\label{n1results}

\begin{figure}
\plotone{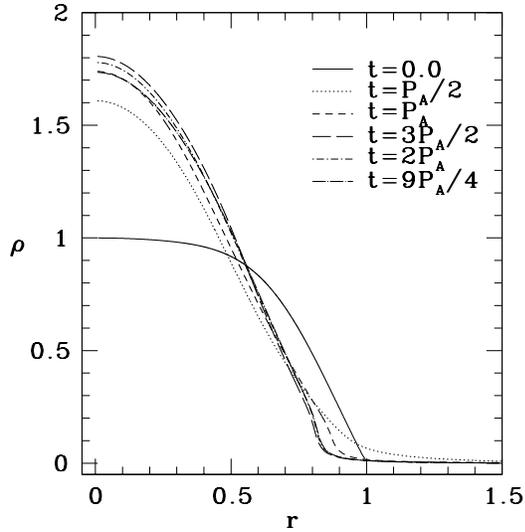}
\caption{Density profiles for Case IV ($n=1$) with zero viscosity 
  ($\eta = 0$). The radius $r$ and density $\rho$ are plotted in nondimensional
  units according to equation (\ref{nondim}).  This plot shows a contraction
  of the inner material with the outer material forming a diffuse atmosphere.
  \label{denprofIVeta0}}
\end{figure}

\begin{figure}
\plotone{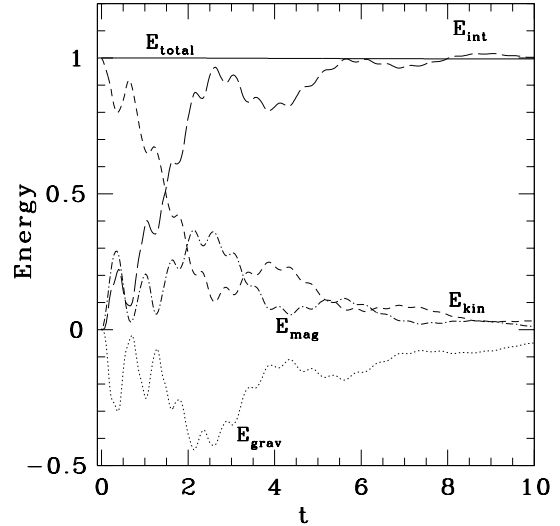}
\caption{Energy evolution for Case IV ($n=1$) with $\eta = 0$.  
  All energies are defined as in Fig.\ \ref{balIeta}.  The sharp 
  increase in $E_{\rm int}$ and sharp decrease in $E_{\rm grav}$ 
  correspond to the contraction of the interior shells.  Then 
  as the outer layers continue to expand, $E_{\rm grav}$  
  increases. Also, $E_{\rm mag}$ grows to an appreciable 
  fraction of the initial kinetic energy before approaching 
  zero.  Note the oscillations in the energy components.
  The sum of the energies is conserved and is plotted as the 
  horizontal solid~line. \label{balIVeta0}}
\end{figure}

\begin{figure}
\plotone{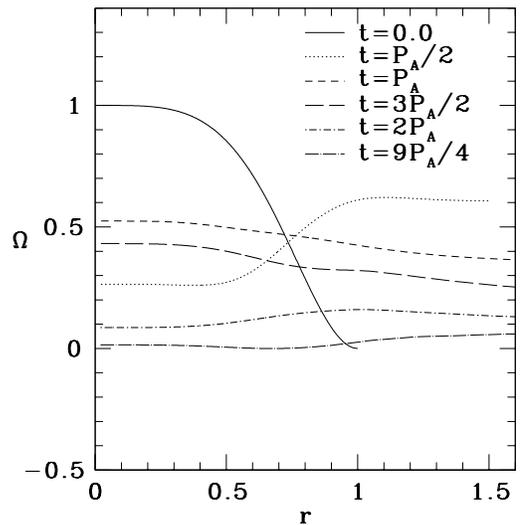}
\caption{Angular velocity profiles for Case IV ($n=1$) with 
  $\eta = 0.2$.  The radius $r$ and angular velocity $\Omega$ are 
  plotted in nondimensional units according to equation (\ref{nondim}).  
  Viscosity and magnetic braking drive the star to uniform~rotation.
  \label{omprofIVeta}}
\end{figure}

We now consider the differentially rotating $n=1$ case (Case IV) 
without viscosity.  This case has $\beta = 0.168$, which is below the
upper limit for uniform rotation, $\beta_{\rm max}$.  Recall that
all differentially rotating $n = 1$ models built according to 
equation (\ref{rotlaw}) have $\beta < \beta_{\rm max}$. However, 
$\beta = 0.168$ is near the upper limit of $\beta$ for the given 
rotation law. Even though this case can relax to uniform rotation 
without appreciable restructuring, we find that this does not happen.  
The star contracts slightly so that the central density grows by a 
factor of $1.819$ and sets up a large diffuse atmosphere (see 
Fig.\ \ref{denprofIVeta0}).  Figure~\ref{balIVeta0} shows the
evolution of the different contributions to the energy.  We find 
that the energy in magnetic fields increases to a significant fraction 
of the initial kinetic energy, but most of the initial kinetic energy 
is ultimately converted into internal energy through shock heating.  
It appears that the final state will be a hot, uniformly rotating star 
with a dense core and a diffuse atmosphere.  We can also see 
oscillations on the energy plot with a period of 
$\sim 0.76\, P_A$.  This is close to the Alfv\'en period, $P_{\rm A}$, 
as defined in equation (\ref{PA}).  The dynamical timescale for this 
evolution is $t_{\rm dyn}=0.111$, which is much shorter than the 
Alfv\'en time.  The virial diagnostic shows that this evolution is 
quasistatic and similar to the $n=3$ cases.  We now consider the 
effects of adding a shear viscosity.  As 
before, the cylinder develops a large diffuse atmosphere and contracts, 
but does so slightly less severely than in the case with $\eta = 0$.  
Radial oscillations occur again with roughly the same period 
($P \approx 0.76\, P_A$), but these oscillations are damped after a 
few periods due to the shear viscosity.  In the $n = 1$ inviscid case, 
the oscillations make determining the final state difficult, but, 
as can be seen in Fig.\ \ref{omprofIVeta}, the cylindrical model
with viscosity tends to uniform rotation by a time $t \sim P_A$.


\subsection{\em The ${\mathbf n=5}$ Results}
\label{n5results}


\begin{figure}
\plotone{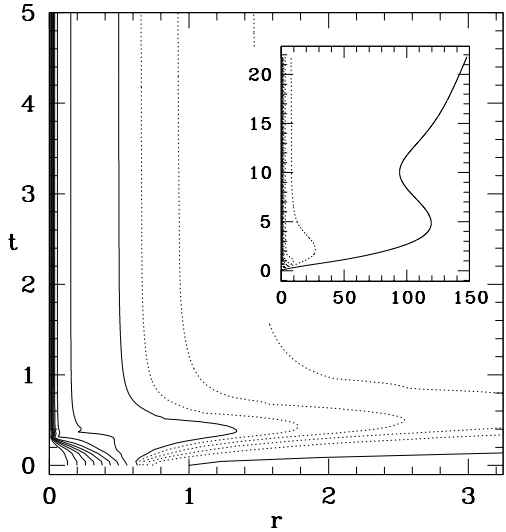}
\caption{Spacetime diagram featuring Lagrangian mass tracers
  for Case VI ($n=5$) without viscosity ($\eta = 0.0$). In the large plot, 
  the solid lines enclose mass fractions $0.10,0.20,\dots,1.0$ while the 
  dotted lines enclose fractions $0.92, 0.94, 0.96$, and $0.98$.  In the 
  inset, the solid lines have the same meaning, while the dotted lines 
  enclose fractions $0.91, 0.91$ \dots , $0.99$.  The outermost line in the 
  inset encloses a mass fraction of $1.0$.  The growth of the azimuthal 
  magnetic field causes a severe contraction of the inner shells 
  and large ejection of the outer layers.  The inset shows that the 
  outer layers form a diffuse~atmosphere. \label{tracerVIeta0}}
\end{figure}

We now describe results for the slowly differentially rotating $n=5$ 
case (Case V) without viscosity.  For this case, we choose 
$\beta = 0.0183 < \beta_{\rm max}$.  Therefore, we do not 
expect strong radial motions in this case.  This case behaves much like 
the analogous $n=3$ case (Case II), so we do not show any plots for it.  
The major difference is that this cylinder experienced a slight 
contraction and radial oscillations which increase in magnitude 
toward the surface.  The central density grows only by a factor of
$1.170$ over three Alfv\'en periods.  We also see oscillations in 
the magnetic field with a period $\sim 0.195\, P_A$, much shorter than 
the Alfv\'en period.  The dynamical timescale for this evolution
is $t_{\rm dyn}=0.0020$, which is again much shorter than the Alfv\'en 
time.  As in the $n=3$ cases, the virial diagnostic shows that this 
evolution is quasistatic.  The only mechanism to dissipate energy is 
through shock heating, and, since the behavior is mild, this would 
require a very long run to reach a final state.  Adding a significant
shear viscosity to this model results in damping of the radial 
oscillations and rapid approach to uniform rotation, as in Case~II.


\begin{figure}
\plotone{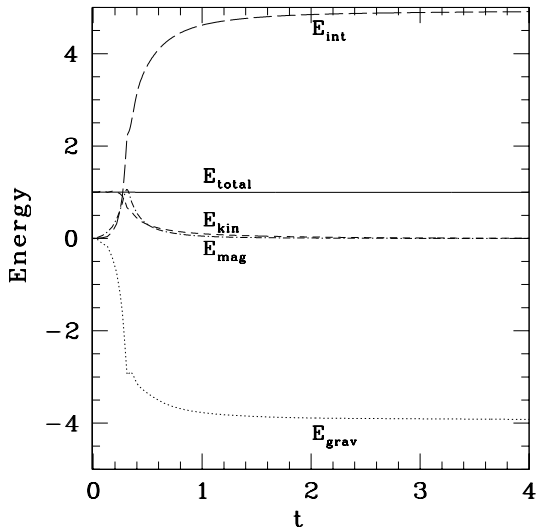}
\caption{Energy evolution for Case VI ($n=5$) with 
  $\eta = 0$. All energies are defined as in Fig.\ \ref{balIeta}.  
  The sharp increase in $E_{\rm int}$ and sharp decrease in $E_{\rm grav}$ 
  again correspond to the contraction of the interior shells.  Also, 
  $E_{\rm mag}$ grows to an appreciable fraction of the initial kinetic 
  energy before approaching~zero. \label{balVIeta0}}
\end{figure}

\begin{figure}
\plotone{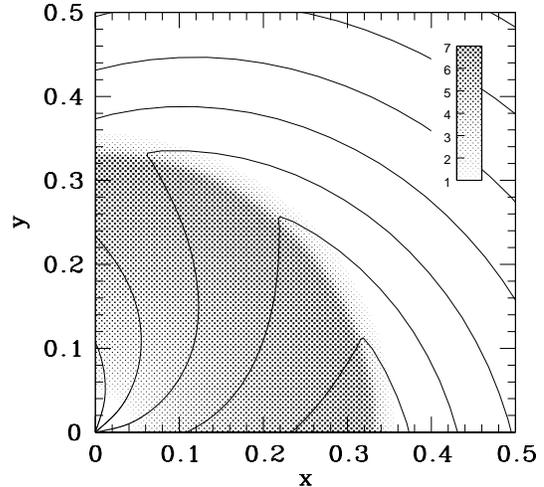}
\caption{Outgoing shock for Case VI with $\eta = 0$ at time
  $t = 0.24\,P_{\rm A} = 0.40$.  (Only one quadrant is shown.)  The shock
  front encloses $\sim 83 \%$ of the total mass.  The ratio
  $[P/\rho^{\Gamma}]/[P/\rho^{\Gamma}]_{t=0}$ is given by the gray shading,
  with magnitudes shown in the bar on the upper right.  The solid lines
  are representative magnetic field lines, calculated as in Fig.\ 
  \ref{twistIeta0}.  The discontinuity indicates a change in the sign 
  of~$B_{\phi}$. \label{n5shock}}
\end{figure}

\begin{figure}
\plotone{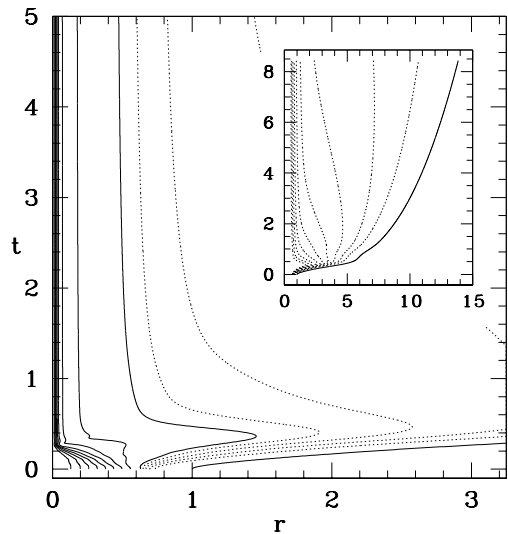}
\caption{Spacetime diagram featuring Lagrangian mass tracers
  for Case VI ($n=5$) with viscosity ($\eta = 0.2$). The lines have the 
  same meaning as in Fig.\ \ref{tracerVIeta0}.  Notice that the 
  contraction is slightly less severe and the ejection much less 
  severe than the case without~viscosity. \label{tracerVIeta}}
\end{figure}

Next, we examine the results for the rapidly differentially rotating 
$n=5$ case (Case VI) with $\eta = 0$ and where 
$\beta = 0.361 \gg \beta_{\rm max}$.  The motion of the Lagrangian mass 
tracers for this case is shown in Fig.\ \ref{tracerVIeta0}.  The 
inner $\sim 91\% $ of the mass undergoes a severe contraction.  The inset 
figure shows that the outer $\sim 9\% $ of the mass expands to large radii, 
forming a diffuse atmosphere.  The time evolution of the various 
components of the energy is shown in Fig.\ \ref{balVIeta0}, which clearly 
shows the strong contraction.  As in the $n=3$ cases, we see that magnetic 
braking leads to significant changes in the structure when the rotation is 
strong.  The final state for Case VI with $\eta = 0$ has 
$\langle P/\rho^{\Gamma} \rangle = 2.65$, which indicates that the strong  
contraction resulted in significant shock heating as in Case III. 
Figure~\ref{n5shock} gives a snapshot of an outgoing shock propagating
through the envelope in the Case IV evolution.  The grayscale shows
$[P/\rho^{\Gamma}]/[P/\rho^{\Gamma}]_{t=0}$, a local measure of heating.  
The solid lines represent magnetic field lines, which are clearly 
discontinuous at the shock front. However, examining the virial again 
reveals that this evolution is nearly quasistatic.
These results can be compared with the analogous case with significant
shear viscosity present ($\eta = 0.2$).  Figure~\ref{tracerVIeta} shows the 
mass tracers for this case.  A contraction similar to that seen in Case VI 
with $\eta = 0$ is seen again here. However, the inset plot shows that the 
outer shells are not ejected to such large distances as they are in the 
previous case.  Also, the central density increases by a factor of $135.9$ 
in this case, but $153.39$ in the case without significant shear viscosity.  
One may conclude that the presence of viscosity has again resulted in 
slightly milder~behavior.

\begin{table*}
\begin{center}
\centerline{\sc Table 3}
\centerline{\sc Summary of Results}
\vskip 6pt
\begin{tabular}{c c c c c c c c c}
\hline
\hline
\multicolumn{4}{c}{Initial Data Characteristics} & Dynamical Behavior & 
\multicolumn{4}{c}{Final State Characteristics\tablenotemark{a}} \\
Case & $\beta$ & $n$ & $\eta$ &  & 
$\langle P/\rho^{\Gamma} \rangle$ & 
$(R/R_0)_{50}$ & $(R/R_0)_{90}$ & $\rho_c/\rho_c(0)$ \\
\hline
 I   & 0.043 & 0.001 & 0.0   & \small{standing Alfv\'en wave}            & 1.00 & 1.00 & 1.00 & 1.00  \\*
     &       &       & 0.2   & \small{damped oscillations}               & \nodata & 1.02 & 1.08 & 1.00  \\ \hline
 II  & 0.030 & 3.0   & 0.0   & \small{expansion, oscillations}           & 1.00 & 0.99 & 1.00 & 1.02  \\*
     &       &       & 0.2   & \small{slight expansion}                  & 1.00 & 0.99 & 1.00 & 1.02  \\*
 III & 0.191 & 3.0   & 0.0   & \small{core contraction}                  & 1.11 & 0.52 & 0.57 & 3.71  \\*
     &       &       & 0.2   & \small{core contraction}                  & 1.40 & 0.53 & 0.63 & 3.50  \\ \hline
 IV  & 0.168 & 1.0   & 0.0   & \small{slight contraction}                & 1.07 & 0.80 & 0.88 & 1.82  \\*
     &       &       & 0.2   & \small{slight contraction}                & 1.22 & 0.82 & 0.95 & 1.77  \\ \hline
 V   & 0.018 & 5.0   & 0.0   & \small{slight contraction, oscillations}  & 1.00 & 0.92 & 0.91 & 1.17  \\*
     &       &       & 0.2   & \small{slight contraction}                & 1.00 & 0.92 & 0.91 & 1.17  \\*
 VI  & 0.361 & 5.0   & 0.0   & \small{strong core contraction}           & 2.65 & 0.08 & 0.78 & 153.4 \\*
     &       &       & 0.2   & \small{strong core contraction}           & 2.54 & 0.09 & 0.73 & 135.9 \\
\hline
\hline
\end{tabular}
\vskip 12pt 
\begin{minipage}{16cm}
${}^{a}$ {$\langle P/\rho^{\Gamma} \rangle$ in units of
$\langle P/\rho^{\Gamma} \rangle_{t=0}$; values larger than 1
indicate heating. $(R/R_0)_i$ denotes the radius containing
a fraction $i$ of the mass, in units of its initial value.}
\end{minipage}
\end{center}
\end{table*}

\section{CONCLUSIONS}
\label{conclusions}

Poloidal seed magnetic fields and viscosity have important dynamical
effects on differentially rotating neutron stars.  Even for a small
seed magnetic field, differential rotation generates toroidal Alfv\'en
waves which amplify and drive the star toward uniform rotation.  This 
magnetic braking process can remove a significant amount of rotational 
energy from the star and store it in the azimuthal magnetic field.  
Though it acts on a longer timescale, shear viscosity also drives the 
star toward uniform rotation.  For a hypermassive star supported against 
collapse by differential rotation, magnetic braking and viscosity 
can lead ultimately to catastrophic~collapse.

The strength of the differential rotation, the degree of 
compressibility, and the amount of shear viscosity all affect the
response of differentially rotating cylinders to the initial magnetic
field.  Our calculations have shown very different behavior when 
$\beta = T/|W|$ is below the upper limit for uniform rotation, 
$\beta_{\rm max}$, than when it is above this limit.  Simulations 
for $n=3$ and $n=5$ with $\beta < \beta_{\rm max}$ show that the 
cylinders oscillate and either expand or contract slightly to 
accommodate the effects of magnetic braking and viscous damping.  
Large changes in the structure are not seen for these cases.  For 
$\beta > \beta_{\rm max}$, on the other hand, the outer layers are 
ejected to large radii, while most of the star contracts 
quasistatically.  In a simulation of a relativistic hypermassive star 
with more realistic geometry, this behavior would likely correspond 
to quasistatic contraction leading to catastrophic collapse and escape 
of some ejected material.  The dynamical behavior is more extreme for 
models with greater compressibility.  This result is reasonable since 
softer equations of state allow stronger radial motions.  Inclusion 
of a shear viscosity often moderates the behavior of an evolving model, 
pacing angular momentum transport and damping the toroidal Alfv\'en 
waves which arise due to differential rotation.  In particular, the 
numerical simulations in the rapidly rotating $n=3$ and $n=5$ cases 
show that the contraction of the core and ejection of the outer 
shells is milder when a significant shear viscosity is present.  Our 
results are summarized in Table~3.

\newpage

Though our model for differentially rotating neutron stars is highly 
idealized, it accommodates magnetic fields, differential rotation,
viscosity, and shocks in a simple, one-dimensional Lagrangian MHD
scheme.  In addition, we are able to handle the wide disparity between
the dynamical timescale and the Alfv\'en and viscous timescales.  
This disparity will likely prove taxing for relativistic MHD 
evolution codes since it will be necessary to evolve the system for
many dynamical timescales in order to see the effects of the magnetic
fields.  Because the cylindrical geometry of the present calculation
greatly simplifies the computational problem, this class of numerical 
models is useful in developing intuition for the physical effects 
which almost certainly play an important role in nascent neutron 
stars.  More realistic evolutionary calculations of magnetic braking 
in neutron stars will serve to further crystallize our understanding 
of the evolution of differentially rotating neutron stars, including 
hypermassive stars that may arise through binary merger or core collapse, 
and may shed light on the physical origins of gravitational wave 
sources and gamma ray bursts.  Several additional physical effects are 
expected to be important.  For example, Shapiro (Paper I) showed that 
the presence of an atmosphere outside of the star allows efficient 
angular momentum loss by partial transmission of Alfv\'en waves at 
the surface.  This will have important effects on the dynamical 
behavior of a neutron star relaxing to uniform rotation.  In addition, 
our treatment does not account for evolution of the magnetic field due 
to turbulence and convection.  However, even without these additional 
physical processes, our calculations reveal qualitatively important 
features of the effects of magnetic braking on the stability of 
neutron stars.  We hope to continue to pursue these questions through 
more realistic computational investigations, including the effects 
of general~relativity.

\acknowledgements
It is a pleasure to thank Charles Gammie, Yuk Tung Liu, and Pedro 
Marronetti for useful discussions.  This work was supported in part 
by NSF Grants PHY-0090310 and PHY-0205155 and NASA Grant NAC 5-10781 
at~UIUC.

\appendix

\section{THE VIRIAL THEOREM}
\label{virial}

In this appendix, we closely follow the derivation of the virial
theorem in Shapiro \& Teukolsky (1983) Section 7.1.  We will discuss 
the inviscid case, because we do not expect viscosity to play a dominant 
role in equilibrium stars.  We start by dotting the position vector 
$\mathbf{x}$ into equation (\ref{navier}). We then multiply through 
by $\rho$, integrate over a cylindrical volume of length $L$ and radius 
$R$, and divide by $L$ (we will work with quantities per unit length 
as we did in section 3.3).  Define $T$, $\Pi$, and $W$ as in section 
3.3 and the moment of inertia per unit length~as
\begin{equation}
I \equiv \int_0^R \rho r^2 d\mathcal{A}.  \label{Idef}
\end{equation}
The left hand side of equation (\ref{navier}) then~becomes
\begin{equation}
\frac{1}{2}\frac{d^2I}{dt^2}-2T.
\end{equation}
The gravitational term is $W$.  We will calculate the pressure term in 
detail to display a feature of this geometry.  The pressure term~is
\begin{eqnarray}
-\frac{1}{L}\int \mathbf{x}\cdot \mathbf{\nabla}Pd^3x &=&
-\frac{1}{L}\int \mathbf{\nabla}\cdot (P\mathbf{x})d^3x  
+\frac{1}{L}\int P\mathbf{\nabla}\cdot (\mathbf{x})d^3x \nonumber \\
&=&-\Pi +3 \Pi,
\end{eqnarray}
where we have used 
\begin{eqnarray}
-\frac{1}{L}\int \mathbf{\nabla}\cdot (P\mathbf{x})d^3x &=&
-\frac{1}{L}\oint P\mathbf{x}\cdot \hat{\mathbf{n}}d\mathcal{A}  \nonumber \\
&=&-\Pi.
\label{pieqn}
\end{eqnarray}
The difference between this infinite cylinder case and the case for
bounded distributions is that we must keep a term from the endcaps 
when applying the divergence theorem in equation (\ref{pieqn}).
Applying similar manipulations to the magnetic terms shows that these 
terms cancel.  Collecting these results~gives
\begin{equation}
\frac{1}{2}\frac{d^2I}{dt^2} = 2T+W+2\Pi.  \label{Virial}
\end{equation}
Assuming steady-state ($d/dt = 0$) gives the virial theorem quoted 
in equation~(\ref{vir}).

\section{STABILITY OF CYLINDRICAL POLYTROPES}
\label{stability}
In this section, we derive by means of an energy variational 
calculation the critical polytropic parameter $\Gamma_{\rm crit}$ 
for the stability of a slowly rotating cylinder against 
radial perturbations, adapting the treatment of stability for 
ordinary stars given in Shapiro and Teukolsky (1983). Because 
we are considering equilibrium models, there is no azimuthal 
magnetic field, and we can take the total energy~as
\begin{equation}
E = E_{\rm int} + T + E_{\rm grav},
\end{equation}
where $T$ is the rotational kinetic energy.  Consider a 
sequence of rotating equilibrium cylinders with constant 
angular momentum per unit length, $J$, parameterized by central 
density, $\rho_c$.  The equilibrium mass per unit length 
$\mu_t$ is determined by the condition 
$\partial E/\partial\rho_c=0$.  Stability then requires 
$\partial^2 E/\partial \rho_c^2 \geq 0$. Using the 
$\Gamma$-law EOS for~polytropes,
\begin{equation}
E_{\rm int} = \int d\mu \frac{K\rho^{\Gamma-1}}{\Gamma-1} = 
k_1 K \mu_t \rho_c^{\Gamma-1},
\label{einternal}
\end{equation}
where $k_1$ is some nondimensional constant of order unity that 
depends on the density profile through $\Gamma$ for slowly rotating
configurations.  The kinetic energy is just $T= J^2/2I$, where $I$ 
is the moment of inertia per unit length.  Since 
$I \propto \mu_t R^2 \propto \mu_t^2 /\rho_c$, one may~write
\begin{equation}
T = k_2 \frac{J^2 \rho_c}{\mu_t^2},
\label{tdefn}
\end{equation}
where $k_2$ is another constant. (We note that $k_1$ and $k_2$ may 
be derived analytically for nonrotating cylinders by rewriting the 
integral in terms of Lane-Emden~variables.)  

Next consider the gravitational potential energy.  As in 
equation~(\ref{energies}), the gravitational energy per
unit length~is 
\begin{equation}
E_{\rm grav} = \onehalf \int dr 2\pi r \rho \Phi 
= \onehalf \int_0^R dr \frac{\partial \mu}{\partial r} \Phi.
\label{egrav1}
\end{equation}
Now from equation (\ref{gradphi}), $\partial \Phi/\partial r = 
2\mu(r)/r$.  Integrating this equation inward from a fiducial
radius $r_0$ outside of the cylinder~gives
\begin{equation}
\Phi(r) = 2\mu_t \ln (r/r_0), \,\,\,\,\,  (r \geq R).
\end{equation}
We now integrate the right-hand side of equation (\ref{egrav1}) by
parts, insert the forms for $\Phi$ and its radial derivative, 
and~obtain
\begin{equation}
E_{\rm grav} = \mu_t^2 \ln(R/r_0) - \int dr \frac{\mu^2(r)}{r}.
\label{egrav2}
\end{equation}
Any terms independent of $\rho_c$ may be discarded for the 
purposes of this variational calculation.  The second term on 
the r.\  h.\  s.\  of equation~(\ref{egrav2}) scales as $\mu_t^2$ and is 
independent of $\rho_c$.  However, the first term depends on 
$\rho_c$ since $R^2 \propto \mu_t/\rho_c$.  Then 
\begin{equation}
E_{\rm grav} = \onehalf \mu_t^2 \ln\left(\frac{\mu_t}{\rho_c r_0^2}
\right) + {\rm constants}.
\label{egrav3}
\end{equation}

From equations (\ref{einternal}), (\ref{tdefn}), and (\ref{egrav3}), 
the total energy is (up to additive~constants)
\begin{equation}
E = k_1 \mu_t K\rho_c^{\Gamma-1} + k_2\frac{J^2\rho_c}{\mu_t^2} 
+\onehalf \mu_t^2 \ln\left(\frac{\mu_t}{\rho_c r_0^2}\right).
\end{equation}
The conditions for equilibrium and for the onset of instability~become
\begin{eqnarray}
\frac{\partial E}{\partial \rho_c} = & 0 & = k_1 (\Gamma-1) \mu_t K
\rho_c^{\Gamma-2} + k_2\frac{J^2}{\mu_t^2}-\frac{\mu_t^2}{2\rho_c} 
\nonumber \\
\frac{\partial^2 E}{\partial \rho_c^2} = & 0 & = 
k_1 (\Gamma-1)(\Gamma-2)\mu_t K \rho_c^{\Gamma-3} + 
\frac{\mu_t^2}{2\rho_c^2}.
\end{eqnarray}
The first equation is equivalent to the condition for hydrostatic
equilibrium (see eq.\ [\ref{Hydrostead}]).  Inserting the first 
equation into the second to eliminate the term proportional to 
$k_1$ gives, after~rearranging,
\begin{equation}
\onehalf(\Gamma-1) - (\Gamma-2)\frac{k_2 J^2\rho_c}{\mu_t^2}
\label{gammainter}
\frac{1}{\mu_t^2} = 0.
\end{equation}
Now, by equation (\ref{Wint}), $\mu_t^2 = |W|$, where $W$ is the gravitational
term appearing in the virial theorem.  We also use equation~(\ref{tdefn})
to identify the kinetic energy $T$. Since equation~(\ref{gammainter}) 
is the condition marking the onset of instability, solving for $\Gamma$
gives the critical~value: 
\begin{equation}
\Gamma_{\rm crit} = \frac{(1-4T/|W|)}{(1-2T/|W|)} \;\;\;\; ({\rm cylinders}).
\end{equation}
By contrast, the result for ordinary bounded stars is (Shapiro \& Teukolsky~1983)
\begin{equation}
\Gamma_{\rm crit} = \frac{4}{3}\frac{(1-5T/2|W|)}{(1-2T/|W|)}
\;\;\;\; ({\rm bounded~stars}).
\end{equation}
Note that, for nonrotating cylinders with $T/|W|=0$, the critical
polytropic parameter is $\Gamma_{\rm crit} = 1$.  For larger values
of $T/|W|$, $\Gamma_{\rm crit}$ decreases.  But since $\Gamma = 1$ is
the smallest $\Gamma$ obtainable for any positive polytropic index $n$,
all equilibrium cylindrical polytropes are radially~stable. 

\section{NUMERICAL TECHNIQUES}
\label{numerical}

We solve the MHD evolution equations using a one-dimensional Lagrangian
finite differencing scheme in which the cylinder is partitioned in 
the radial direction into $N$ shells of equal mass per unit length.  
The evolution equations are differenced on staggered meshes in time and
space.  The quantities $r$, $\upsilon_r$, $j$, and $\mu$ are defined 
on the shell boundaries while $P$, $V \equiv 1/\rho$, $\varepsilon$,
$B_{\phi}$, and $q$, the artificial viscosity, are defined at the shell
centers.  Integral and half-integral spatial indices correspond to 
shell boundaries and shell centers, respectively.  For example, one has
$r_i$ where $i=1, 2, \dots, N+1$ and $P_{i+1/2}$ where $i = 1,2,\dots, N$.
Updating at each timestep occurs in leap-frog fashion, with $\upsilon_r$,
$q$, and $j$ defined at times with half-integral indices ($t^{n+1/2}$) 
and $r$, $P$, $V$, $B_{\phi}$, and $\varepsilon$ defined at times~$t^n$.

Shocks are handled using artificial viscosity as described by 
Richtmyer and Morton (1967).  In physical units, the artificial 
viscosity prescription~is
\begin{equation}
q = \cases{
\frac{(a\delta r)^2}{V}\left(\frac{\partial\upsilon_r}{\partial r}
\right)^2, &if $\partial\upsilon_r/\partial r < 0$;\cr
      0, &otherwise.\cr}
\end{equation}
where $\delta r$ is the typical spatial mesh size and the thickness 
of the resulting smoothed shock is approximately $a^2 \delta r$ (we usually 
take $a = 1.75$). The timestep is governed by the usual Courant 
condition in the absence of viscosity: $\delta t =  b \delta r/c_s$, 
where $c_s$ is the local sound speed and $b$ is an constant 
less than unity (we usually take $b = 0.3$).  If shear viscosity is 
present, the condition~becomes 
\begin{equation}
\delta t = {\rm min} \left\{ b\frac{\delta r}{c_s},
b\frac{(\delta r)^2}{2D_{\rm eff}}\right\},
\end{equation} 
where $D_{\rm eff} = 4\eta/3\rho$ is an effective diffusion~coefficient.

We found it convenient to use a different nondimensionalization scheme
for our numerical scheme than was employed in the discussion above 
(eq.\  [\ref{nondim}]).  We will now use the following~definitions: 
\begin{eqnarray}
\label{nondim2}
& & \tilde{r} = r/R,\;\;\; \tilde{t} = t/(R/c_s(0)),\;\;\;
\tilde{\eta} = \eta/(\rho_c c_s(0) R),
\\ \nonumber
& & \tilde{j} = j/(c_s(0) R), \;\;\; \tilde{B}_\phi =
B_\phi/B_0, \\ \nonumber
& & \tilde{\rho} = \rho/\rho_c, \;\;\;  \tilde{\upsilon}_r = 
\upsilon_r/c_s(0), \;\;\; \tilde{\varepsilon} = 
\varepsilon/[K\rho_c^{\Gamma -1}/(\Gamma-1)] \\ \nonumber
& & \tilde{\mu} = \mu/(\rho_c R^2),\;\;\; \tilde{P} =
P/(K\rho_c^{\Gamma}), 
\end{eqnarray}
where $c_s(0)=(\Gamma K\rho_c^{\Gamma-1})^{1/2}$ is the sound 
speed at $r=0$ and $t=0$.  This differs from the previous scheme 
in that it relies on the sound speed as a velocity 
scale rather than the Alfv\'en speed.  We finite-difference the 
evolution equations (\ref{vrevolve1})-(\ref{eintevolve1}), using 
equations (\ref{nondim2}) to write the system in nondimensional
form.  (Below we drop the tildes on all nondimensional quantities.)  
If no shear viscosity is present, all derivatives are approximated 
by second order centered differences, so that the scheme is second 
order in space and time (except at the boundaries).  The terms 
which account for shear viscosity are not time centered ({\em i.e.}\ 
these terms are {\em first} order in time).  This simplification 
does not significantly affect our results since the viscous terms 
are small compared to the other terms in the evolution~equations. 

Before giving the evolution system, we will introduce several 
definitions and conventions to simplify the notation.  First, 
we will drop the $r$ index on $\upsilon_r$, the $\phi$ index on 
$B_{\phi}$, and the tildes.  We also define 
$\theta \equiv \upsilon_{\rm A}/c_s(0)$.  In all of the following
equations, the operator $\Delta$, which takes a spatial difference,
will be defined by $\Delta Q_i = Q_i - Q_{i-1}$.  In some of the 
equations, it is useful to define space and time averages for grid 
positions with half-integral indices.  For example, 
$r_{i+1/2}^n \equiv (r_{i+1}^n+r_i^n)/2$ and $r_i^{n+1/2} \equiv 
(r_i^{n+1}+r_i^n)/2$.  Then the evolution equation for specific 
angular momentum (eq.~[\ref{jevolve1}]) becomes
\begin{equation}
\frac{j_i^{n+1/2}-j_i^{n-1/2}}{\delta t}  =  
\frac{2\pi\theta^2}{\delta\mu}\Delta(r_{i+1/2}^n B_{i+1/2}^n)
 + \frac{2\pi\eta}{\delta\mu}\Delta \left[\frac{1}{2}\left(
\frac{(r_{i+1}^n)^3+(r_i^n)^3}{\Delta r_{i+1}^n}\right)
\Delta\left(\frac{j_{i+1}^{n-1/2}}{(r_{i+1}^n)^2}
\right)\right].  \label{jupdate}
\end{equation}
We have used the fact that the Lagrangian mass increment between two 
shells, $\delta\mu$ is a constant.  The radial velocity is updated 
next as~follows:
\begin{eqnarray}
\frac{\upsilon_i^{n+1/2}-\upsilon_i^{n-1/2}}{\delta t} & = & 
-\frac{2\pi r_i^n}{\Gamma\delta\mu}\Delta(P_{i+1/2}^n) - 
\left(\frac{n\xi_1^2}{2\pi}\right)\frac{\mu_i}{r_i^n}
-\frac{\pi\theta^2}{\delta\mu r_i^n}\Delta((r_{i+1/2}^n)^2 B_{i+1/2}^n) 
\nonumber \\
& & + \frac{1}{2(r_i^n)^3}[(j_i^{n+1/2})^2 + (j_i^{n-1/2})^2] \nonumber \\
& & + \frac{8\pi\eta}{3\delta\mu r_i^n}\Delta\left[ \frac{1}{2}\left(
\frac{(r_{i+1}^n)^3+(r_i^n)^3}{\Delta r_{i+1}^n}\right)\Delta\left(
\frac{\upsilon_{i+1}^{n-1/2}}{r_{i+1}^n}\right) \right]. 
\end{eqnarray}
The $\xi_1$ appearing in the gravitational potential term of the above
equation is just the nondimensional surface radius from the 
solution of the Lane-Emden equation.  It arises here only because of our
choice of nondimensionalization.  Next, we update the locations of shell 
boundaries according~to 
\begin{equation}
\frac{r_i^{n+1}-r_i^n}{\delta t} = \upsilon_i^{n+1/2}.
\end{equation}
The specific volume of each shell is then updated to reflect the new
shell boundaries:
\begin{equation}
V_{i+1/2}^{n+1} = V_{i+1/2}^0 \left(\frac{\Delta(r_{i+1}^{n+1})^2}
{\Delta(r_{i+1}^0)^2}\right),
\end{equation}
where the superscript ``0'' refers to the initial value (this equation merely 
expresses the fact that the volume of a thin shell is $dV = \pi Ldr^2$).  
We next integrate the evolution equation for the magnetic~field. 
\begin{equation}
\frac{B_{i+1/2}^{n+1}-B_{i+1/2}^n}{\delta t} = 
-\frac{1}{2}(B_{i+1/2}^{n+1}+B_{i+1/2}^n)\left(\frac{\Delta \upsilon_{i+1}^{n+1/2}}
{\Delta r_{i+1}^{n+1/2}}\right) + \frac{1}{\Delta r_{i+1}^{n+1/2}}\Delta
\left(\frac{j_{i+1}^{n+1/2}}{(r_{i+1}^{n+1/2})^2}\right).
\end{equation}  
This equation can be solved for the updated magnetic field in each cell,
$B_{i+1/2}^{n+1}$.  The next quantity to update is the artificial~viscosity:
\begin{equation}
q_{i+1/2}^{n+1/2} = \cases{
\frac{2 a^2 \Gamma}{V_{i+1/2}^{n+1/2}}
(\Delta\upsilon_{i+1}^{n+1/2})^2, 
& if $\Delta\upsilon_{i+1}^{n+1} < 0$; \cr 
0, & otherwise. \cr }
\end{equation}
The specific internal energy is updated according to the First Law of 
Thermodynamics with entropy generation terms due to the presence of 
viscosity (see eq.\ [\ref{eintevolve2}]):
\begin{equation}
\frac{\varepsilon_{i+1/2}^{n+1}-\varepsilon_{i+1/2}^{n}}{\delta t} = 
-\frac{1}{2n}(P_{i+1/2}^{n+1}+P_{i+1/2}^{n})
\left(\frac{V_{i+1/2}^{n+1}-V_{i+1/2}^n}{\delta t}\right) + 
\frac{\Gamma\eta}{n}V_{j+1/2}^{n+1/2} f_{j+1/2}^{n+1/2}(\upsilon,j,r),
\label{energyupdate}
\end{equation}
where
\begin{eqnarray}
f_{i+1/2}^{n+1/2}(\upsilon,j,r) & \equiv & \frac{2}{9}\left[
\left\{2\frac{\Delta\upsilon_{i+1}^{n+1/2}}{\Delta r_{i+1}^n}
-\frac{\upsilon_{i+1/2}^{n+1/2}}{r_{i+1/2}^n}\right\}^2 
+\left\{\frac{(r_{i+1}^n)^2+(r_i^n)^2}{2\Delta r_{i+1}^n}
\Delta\left(\frac{\upsilon_{i+1}^{n+1/2}}
{(r_{i+1}^n)^2}\right)\right\}^2 \right. \nonumber \\
& & + \left.\left\{2\frac{\Delta (r_{i+1}^n\upsilon_{i+1}^{n+1/2})}
{\Delta(r_{i+1}^n)^2}\right\}^2 \right]
+ \left\{\frac{r_{i+1/2}^n}{\Delta r_{i+1}^n}
\Delta\left(\frac{j_{i+1}^{n+1/2}}{(r_{i+1}^n)^2}\right)\right\}^2.
\end{eqnarray}
Note that the terms due to viscous dissipation are not time-centered.
Equation~(\ref{energyupdate}) may be solved for the updated value 
of the specific internal energy, $\varepsilon_{i+1/2}^{n+1}$,
by first substituting for the (as yet unknown) updated pressure,
$P_{i+1/2}^{n+1}$, according to the $\Gamma$-law~EOS:
\begin{equation}
P_{i+1/2}^{n+1} = \varepsilon_{i+1/2}^{n+1}/V_{i+1/2}^{n+1}.
\label{pupdate}
\end{equation}
Finally, the pressure itself is updated using 
equation~(\ref{pupdate}).  In order to evolve the system from one
timestep to the next, equations (\ref{jupdate})-(\ref{pupdate}) are 
used in the order as given above to updated each quantity. These 
equations only apply in the interior regions of the grid and some 
must be modified to accommodate boundary conditions on the cylinder 
axis ($r=0$) and at the surface.  These modifications result in 
only first order spatial accuracy near these~boundaries.

All of the calculations described in this paper were performed on 
single processors and none lasted more than a few hours.  As an 
example, consider the rapidly rotating $n=3$ cylinder with 
$\eta = 0$. This run required 2400 Lagrangian mass shells and 
each timestep required 7.295 ms of CPU time on a Pentium IV, 
2.0 GHz processor.  The total length of this run was 
$4.4 \times 10^6$ steps.  Our runs required spatial resolution 
in the range of 400-2400 Lagrangian mass~shells.

\end{document}